\documentclass[sn-mathphys,Numbered]{sn-jnl}

\usepackage{verbatim}%
\usepackage{setspace}%
\usepackage{graphicx}%
\usepackage{multirow}%
\usepackage{amsmath,amssymb,amsfonts}%
\usepackage{amsthm}%
\usepackage{mathrsfs}%
\usepackage[title]{appendix}%
\usepackage{xcolor}%
\usepackage{textcomp}%
\usepackage{manyfoot}%
\usepackage{booktabs}%
\usepackage{algorithm}%
\usepackage{algorithmicx}%
\usepackage{algpseudocode}%
\usepackage{listings}%
\usepackage{bigints}

\theoremstyle{thmstyleone}%
\theoremstyle{thmstyletwo}%

\theoremstyle{thmstylethree}%

\begin{document}

\title{\bf The development of the concept of exchange forces in the 1930s: close encounters between Europe and Japan and the birth of nuclear theory}


\author[1]{\fnm{Marco} \sur{Di Mauro}}\email{marco.dimauro@unitn.it}
\author*[2,3]{\fnm{Salvatore} \sur{Esposito}}\email{salvatore.esposito@na.infn.it, salvatore.esposito12@unina.it}
\author[3]{\fnm{Adele} \sur{Naddeo}}\email{anaddeo@na.infn.it}
\equalcont{These authors contributed equally to this work.}

\affil[1 ]{\orgdiv{Department of Physics}, \orgname{University of Trento}, \orgaddress{\street{Via Sommarive}, \city{Povo (TN)}, \postcode{38123}, \country{Italy}}}

\affil*[2]{\orgdiv{Department of Physics ``Ettore Pancini"}, \orgname{University of Naples ``Federico II"}, \orgaddress{\street{Piazzale Tecchio 80}, \city{Naples}, \postcode{80125}, \country{Italy}}}

\affil[3]{\orgdiv{Naples' Unit}, \orgname{I.N.F.N.}, \orgaddress{\street{Complesso Universitario di Monte S. Angelo, Via Cinthia}, \city{Naples}, \postcode{80126}, \country{Italy}}}


\abstract{The onset and the development of the concept of exchange force in quantum physics are historically reconstructed,
starting from Heisenberg's seminal contributions in 1926 and going through the great developments in nuclear physics, which 
allowed the emergence of the idea of force mediating virtual quanta. Although most of such work was performed in Europe, the
last and decisive effort in this long path was carried out by Japanese scientists in the 1930s. This is the main
focus of the present work, which retraces the achievements of Yukawa and Tomonaga, whose results and mutual interactions 
are carefully analyzed and related to those of European physicists.}




\maketitle

\section{Introduction}

The concept of exchange interactions dates back to the second half
of 1920s and is deeply related to the quantum mechanical
description of a system of identical particles. The idea was
conceived by Werner Heisenberg shortly after the final theory of
quantum mechanics had been developed
\cite{Heisenberg1,DuckSudarshan,Heisenberg2} and immediately applied by him to
the hard problem of the spectrum of the helium atom
\cite{Heisenberg3}, which had defied all attempts to make sense of
it within the old quantum theory. Exchange interactions underlie
the successful explanation of a variety of phenomena besides
multi-electron atomic spectra, such as homopolar bonds in quantum
chemistry, ferromagnetism, and electron-atom collisions, although
the interpretation of the exchange mechanism is different in each
case \cite{Carson1,Carson2}. The discovery of the neutron by James
Chadwick in 1932 \cite{Chad1932,Chad1932a} brought about, in the
hands of Heisenberg himself, the application of the concept of
exchange forces to the domain of nuclear physics and, finally, of
quantum field theory \cite{PaisIWB,Brink0}, paving the way to the
understanding of fundamental forces as mediated by virtual
particle exchange. The Heisenberg theory of nuclear structure,
based on an exchange mechanism responsible for the interaction
between protons and neutrons
\cite{HeisenbergEX1,HeisenbergEX2,HeisenbergEX3}, Ettore
Majorana's further refinement of nuclear exchange forces
\cite{SIF,Majorana1} and Enrico Fermi's theory of $\beta$-decay
\cite{Fermi1,Fermi2,Fermi3,Fermi4} are the first fruitful developments of Heisenberg's ideas. Despite none of
them could successfully explain the whole complexity of the
nuclear binding force, they constitute very important
milestone towards that difficult task. These works strongly
affected Yukawa Hideki\footnote{In this paper,
excluding quotes, we adopt the Hepburn system to transliterate
Japanese names and, when referring to Japanese people, we follow
the Japanese use of writing the surname before the first
name.} \cite{Tabibito} as well as Tomonaga Shin'ichir\={o}
\cite{TomonagaSpin}, who already had been exposed to the
principles of the new quantum mechanics, by taking part to a cycle
of lectures given by Heisenberg and Dirac in Japan in 1929 and
promoted by Nishina Yoshio \cite{Konagaya,Ito1,Ito2,Kim1}. It took
a couple of years until Yukawa conceived his key idea of a
mediating virtual meson \cite{Yukawa1935,Brown1,Brown2,Darrigol1}.
In the same period, Tomonaga was working on the range of
proton-neutron interactions in Nishina's laboratory
\cite{NishinaTomonagaTamaki1936} and had fruitful exchanges
with Yukawa.

In this paper, we retrace the above developments, with
particular attention to the contribution given by Japanese
physicists to the modern understanding of fundamental forces in
the 1930s, which is carefully analyzed and related to
previous key results. A clear picture emerges also of the
influence of European scientists in shaping the development of
quantum concepts in Japan. The paper is organized as follows.
Sections 2, 3 and 4 are devoted to a brief but careful
analysis of the key ideas underlying the seminal contributions
provided by Heisenberg, Majorana and Fermi, with emphasis on
conceptual issues. Their results are indeed mandatory steps
for a thorough reconstruction of the role played by Japanese
physicists in building the nuclear force theory, which is the
content of Sections 6 and 7. Before that, a brief account of the
situation of theoretical physics in Japan up to the period of
interest is given in Section 5, with the aim of providing the
general context in which Yukawa's as well as Tomonaga's
contributions grew up and developed. Finally, Section 8 contains
our concluding remarks.

\section{Heisenberg's seminal contributions}

The aim of this Section is to give an account of the key ideas
underlying Heisenberg's theory of exchange forces, which marked
the birth of nuclear theory
\cite{HeisenbergEX1,HeisenbergEX2,HeisenbergEX3}.

\subsection{The introduction and first applications of exchange interactions}

The very roots of Heisenberg's work can be found in his famous paper on quantum resonance \cite{Heisenberg1}, dated June 1926, in which he introduced the peculiar non-classical concept of intrinsic symmetry in a concrete attempt to address many-body systems. Besides the obvious desire of extending his brainchild, matrix mechanics, to situations in which more than one particle were involved, Heisenberg's primary concern was to explain the splitting between singlet and triplet states in the spectrum of the helium atom, whose value was inconsistent with the one expected as a consequence of spin interactions between the two atomic electrons. Another problem was the mismatch between the theoretical number of stationary states of a quantum many-body system (obtained by means of Boltzmann statistics) and the stationary states actually observed in nature. According to Heisenberg, the solution of these puzzling issues depended on the understanding of the relationship between Bose-Einstein statistics, quantum mechanics and Pauli's exclusion principle. In his words:

\begin{quote}
We have found a rather decisive argument that your exclusion of
equivalent orbits is connected with the singlet-triplet separation
[...]. Consider the energy written as a function of the transition
probabilities. Then a large difference results, if one -- at the
energy of H atoms -- has transitions to $1S$, or if, according to
your ban [exclusion principle], one puts them equal to zero. That
is, para- and ortho-[helium] do have different energies,
independently of the interaction between magnets [i.e., the
magnetic moments associated with the spinning electrons] (Postcard
addressed to W. Pauli, 5 May 1926, cited in \cite{Mehra1}, p.
737).
\end{quote}
Indeed, the solution of three-body problems, in particular of the
$H_2^{+}$ molecular ion and of the helium atom, was the stumbling
stone which marked the failure of the old Bohr-Sommerfeld atomic
theory, as already experienced by Heisenberg himself together with
Born in 1923 \cite{BornH1}. It soon became clear that new
developments were needed in order to attack the problem and this
consideration led Heisenberg in 1926, now armed with the new
quantum mechanics, to tackle the many-body problem again, trying
to extend his matrix mechanics to the simplest case of the helium
atom \cite{Heisenberg1, Heisenberg3}. He also considered the
general case of a system composed of $n$ identical components,
coupled by two-body interactions. As a useful toy model,
Heisenberg considered two weakly coupled identical harmonic
oscillators. The term ``resonance'' comes in fact from the
classical behavior of such a system.

Besides pointing out the role of the electron-electron interaction
as crucial for the resonance idea, his conclusions rely strongly
on the application of Bose-Einstein statistics as well as Pauli's
exclusion principle. Then, in a second paper \cite{Heisenbergs},
the relevance of the notion of indistinguishable particles with
respect to the application of Bose-Einstein statistics is
recognized and the problem of the stability of the helium atom
addressed via the concept of exchange energy. It is worth pointing
out that, in the above papers, Heisenberg missed the
incompatibility between Bose-Einstein statistics and Pauli's
exclusion principle, even if a few months earlier Fermi had
already introduced a different quantum statistics of the ideal gas
based on a generalization of Pauli's principle \cite{FermiIG1926}.
The same quantum statistics would have been independently obtained
shortly after also by Paul A. M. Dirac, in his paper on the
quantum mechanical treatment of many particles systems
\cite{Dirac1926QS}: it is now known as Fermi-Dirac statistics.

The idea of exchange forces grew up in the subsequent years,
switching from the atomic physics context to quantum
chemistry\footnote{The idea of exchange forces was introduced in
molecular physics by Walter Heitler and Fritz London
\cite{HeitlerLondon1927} and soon became the basis of the quantum
theory of homopolar chemical bond \cite{noiCB1,noiCB2,noiCB3}.
Heitler and London set up a perturbative approach for the solution
of the Schr\"{o}dinger equation for the hydrogen molecule $H_2$,
by taking as unperturbed eigenfunctions the products of single
particle eigenfunctions corresponding to the configuration in
which one of the two electrons is on one of the two nuclei, and
the remaining electron on the other nucleus. The twofold
degeneracy leads to take two linear
combinations of the previous products as unperturbed eigenfunctions, the first one being symmetric and the
second one antisymmetric, respectively. Due to the perturbation, he symmetric combination
corresponds to a lower energy, hence to the formation of a stable molecule. According to Heitler and
London there is a finite probability for the electron of the first
nucleus to be found around the other nucleus, so that a resonance
phenomenon arises, similar to that already introduced by
Heisenberg for the helium atom. Such a phenomenon (termed in
German {\it Austausch}), allowing electrons to exchange places
around the two different nuclei, is crucial for the explanation of
the homopolar chemical bond: in Heitler and London's words, it is
``a characteristic quantum mechanical effect''.} and
ferromagnetism,\footnote{Heisenberg's theory of ferromagnetism
relies on the forces responsible for magnetic order being
identified with quantum mechanical exchange interactions between
electrons \cite{HeisenbergFerro}. According to Heisenberg, the
same mechanism responsible of level splittings between electrons
in singlet and triplet states of two-electron atoms, in this new
context, gives rise to spin alignment.} as well as electron-atom
collisions.\footnote{Here the exchange of the incoming electron
with one of the atomic electrons produces interference terms
modifying the cross-section \cite{Oppenheimer1928}.}

The application of the idea to nuclear physics dates from 1932,
with Heisenberg's seminal contributions
\cite{HeisenbergEX1,HeisenbergEX2,HeisenbergEX3}, paving the way
for future developments that culminated into the application to
quantum electrodynamics and quantum field theory.\footnote{We refer
to Refs. \cite{Carson1,Carson2} for a careful analysis of the
historical development of the notion of exchange interactions, and
in particular of its modifications when passing from a context to
a different one, the final output of this process being the actual
interpretation of fundamental forces as virtual particle
exchange.}

\subsection{Exchange forces and nuclear physics}

Let us now analyze Heisenberg's important work on nuclear
theory.\footnote{See Refs. \cite{MillerPT1985,MillerBook1994} for a
thorough discussion of the connections between visualization and
intuition and, in particular, of the guiding role that changes in
intuition play in determining a corresponding change in the visual
representation used to describe subatomic phenomena.} As a
previous mandatory step, we shall briefly summarize the state of
art of research in nuclear physics in 1932.

Chadwick's discovery of a neutral particle of mass
comparable with that of the proton, i.e. the neutron
\cite{Chad1932,Chad1932a}, is, without any doubt, one of the most
relevant experimental findings of the early 1930s. Indeed, it
provided the solution of some puzzling issues concerning
the nuclear structure, such as the wrong spin and statistics of
several nuclei, for example the $^{14}N$ nucleus. This last
feature was a direct consequence of the existing model for the
nucleus, as built of protons and electrons, which in the case of
the $^{14}N$ nucleus led to fourteen protons and seven electrons,
thus implying that this nucleus should have half-integer spin and
obey Fermi-Dirac statistics,\footnote{In fact, according to the
Ehrenfest-Oppenheimer rule \cite{EOrule}, any composite, including
nuclei, containing an odd (resp. even) number of fermionic
particles should follow the Fermi-Dirac (resp. Bose-Einstein)
statistics. Notice that at the time the issues of spin and
statistics were in principle separated, as the general
spin-statistics theorem had still to come \cite{PaisIWB}.}
at odds with observations of band spectra of the $N_2^+$
molecule, consistent with spin-1 and Bose statistics for
those nuclei (see \cite{PaisIWB,TomonagaSpin} for a
discussion). In order to address this problem, some strange
properties had to be hypothesized for nuclear electrons, for
instance that they did not contribute to the spin and statistics
of the nucleus because their spins and statistics would get
somehow suppressed, so their behavior would not follow quantum
theory and Dirac's equation.\footnote{See Refs.
\cite{PaisIWB,Gamow1931} for comprehensive accounts of the
problems arising within an electron-proton model of the nucleus.}
Other inconsistencies brought by the hypothesis of nuclear
electrons were the incorrect value of the magnetic moments of
nuclei and the inconsistency of the energies expected for
nuclear electrons from the momentum-position uncertainty relation
due to their confinement in the nucleus, as well as both the average
binding energy for a nuclear particle and the typical kinetic
energies of electrons emitted as $\beta$-rays. Related to
this, there was the well-known problem of the continuous spectrum
of such $\beta$-rays, which seemed to threaten the validity of
energy and momentum conservation. All these problems induced some
scientists, most notably Niels Bohr, to believe that quantum
mechanics, as well as the cherished conservation laws of
energy, momentum and angular momentum, did not work at nuclear
scales, and that somehow the electrons ``lost their
individuality'' when confined to such a small
scale.\footnote{Sometimes, as a (rather weak) argument in
favor of such speculations, the fact was adduced that the classical electron
radius is of the same order of magnitude of the nuclear scale,
which in fact is purely accidental.} Bohr's position is vividly
expressed by Tomonaga as follows:

\begin{quote}
Bohr asserted that this difficulty appeared because quantum
mechanics did not work in the nucleus, and therefore the law of
energy conservation was violated. Bohr said that since an electron
with a Compton wavelength of $10^{-11}$ cm was confined in the
small nucleus with a $10^{-13}$ cm radius, the individuality of
the electron was lost in the nucleus. [...] The bottom line is
that according to Bohr's idea, the interior of a nucleus was a
sanctuary that could not be penetrated by quantum mechanics
(\cite{TomonagaSpin}, pp. 159-160).
\end{quote}
As we will see now, the ``sanctuary'' began instead to be
penetrated by Heisenberg who brought quantum mechanics into the
nucleus.  The starting point of Heisenberg's first paper on
nuclear forces \cite{HeisenbergEX1} is the key assumption that
atomic nuclei are built up with protons and neutrons, with no
electrons. Besides a relevant simplification of the theory,
which now only involves heavy and slow particles and thus
can be formulated using ordinary non-relativistic quantum
mechanics, the correct statistics of nuclei such as $^{14}N$ follows at once from further assuming that the neutron has
spin $\frac{1}{2} \hbar$ and satisfies Fermi statistics. But the
problem with the $\beta$-decay energy spectra remained. To address
this and other issues such as the the interaction of nuclei with
$\gamma$-rays, in all three papers Heisenberg adopted a kind
of ``hybrid'' approach in which the neutron is sometimes
considered as an elementary particle, and sometimes as a composite
of an electron and a proton, despite this being obviously in
contrast with the assumptions made for spin and statistics. As a matter of
fact, Heisenberg was sweeping all the problems of the quantum
mechanical description of a proton-electron nucleus inside the
neutron:

\begin{quote}
The neutron will be taken as an independent fundamental particle
which, however, can split, under favorable conditions, into a
proton and an electron, violating the law of conservation of
energy and momentum (\cite{HeisenbergEX1}, translated in
\cite{Brink0}, p. 145).
\end{quote}
At this stage, Pauli's neutrino hypothesis\footnote{The
proposal of a new neutral particle with a tiny mass in order to
get rid of the problem of energy non-conservation in $\beta$-decay
was put forward for the first time by Pauli in an open
letter addressed to Hans W. Geiger and Lise Meitner at a meeting
in T\"{u}bingen in December 1930 \cite{PauliNeutrino1}, and later
publicly expressed at the American Physical Society meeting held
in Pasadena in June 1931. This circumstance is recalled by Pauli
itself at the Solvay Conference in Brussels in 1933: ``In June
1931, during a conference in Pasadena, I proposed the following
interpretation: the conservation laws hold, the emission of beta
particles occurring together with the emission of a very
penetrating radiation of neutral particles, which has not been
observed yet. The sum of the energies of the beta particle and the
neutral particle (or the neutral particles, since one doesn't know
whether there is one or many) emitted by the nucleus in one
process, will be equal to the energy which corresponds to the
upper limit of the beta spectrum'' (\cite{PauliSolvay1933}, p.
324). A detailed account of Pauli's neutrino hypothesis can be
found in Ref. \cite{PaisIWB}.} had not yet been recognized by
Heisenberg, while $\beta$-decay would have been the subject of a
careful analysis at the end of the paper.

The core of Heisenberg's work is the consideration that a correct
description of the nuclei can be given in terms of quantum
mechanics and, in particular, of the interaction between protons
and neutrons. The nature of this interaction needs to be
established. Two main guidelines were followed: first, the force
must be very strong, in order to bind protons and neutrons within
a region of the order of $10^{-13}$ cm, i.e., the nuclear radius,
as can be inferred also from proton-nuclei scattering experiments,
which also pointed at this force being of very short range;
second, experiments indicated a saturation property of nuclear
forces, namely that the binding energy of the nuclei is
proportional to the mass number $A$ (i.e. the  number of particles
inside the nucleus), rather than $A^2$ as would be the case
for a force in which each nucleon interacted with all other
nucleons. The conclusion was that the required interaction
could not be of electromagnetic origin, and should be
analogous to the one responsible of the existence and
stability of the $H_2^+$-ion within the domain of molecular
physics. The proton and the neutron were seen as analogues
of a hydrogen nucleus and a hydrogen atom, and the interaction
could be seen, as its molecular-theoretic counterpart, as due to
the exchange of an electron between the two. Notice that this
picture requires the mentioned problematic composite picture of
the neutron. In Heisenberg's words:

\begin{quote}
If one brings a neutron and a proton to within a distance
comparable to the dimensions of the nucleus, then -- in analogy with
the $H_2^+$ ion -- a change of place [Platzwechsel] of the negative
charge will occur with frequency $\frac{1}{h} J(r)$. The quantity
$J(r)$ corresponds to the Austausch- or more correctly the
Platzwechsel-integral of molecular theory. One can illustrate this
Platzwechsel again with the picture of electrons that have no spin
and obey Bose statistics. But it is probably more correct to
regard the exchange integral $J(r)$ as a fundamental property of
the proton-neutron pair, without wanting to reduce it to motions
of electrons (\cite{HeisenbergEX1}, p. 2; translated in
\cite{Carson2}, p. 104).
\end{quote}
Notice that Heisenberg speaks here about an electron without spin
obeying Bose statistics, in this way preserving the
fermionic nature and the half-integer spin of the neutron. Within
such a picture, the mechanism of the interaction between the
neutron and the proton can be viewed as an exchange of an electron
between the neutron and the proton: the neutron which loses its
electron turns into a proton while the proton, which captures the
electron, becomes a neutron.\footnote{This interpretation of
Heisenberg's exchange forces has been discussed in detail by A. I.
Miller \cite{MillerPT1985,MillerBook1994}.} However he will end
up by hinting to a different interpretation of the exchange
integral $J(r)$, without any reference to moving electrons,
but as related to the exchange of the proton and the neutron
themselves.\footnote{As can be seen from the above quote,
Heisenberg underlines this conceptual shift using a linguistic
subtlety, that has been lost in some translations from German to
English (such as that in \cite{Brink0}). Indeed, Heisenberg
(\cite{HeisenbergEX1}, p. 2) first referred to the exchange
integral as {\it Platzwechsel}, which literally means
``change of position'', rather than as {\it Austausch}, which
instead is adopted within a quantum chemistry context in the case
of systems with two or more identical particles, such as the
hydrogen molecule with two electrons. Indeed, as said, the case of
neutron-proton interaction is the nuclear analogue of the hydrogen
molecular ion, with only one electron moving from one nucleus to
the other and vice versa. Hence in such cases the
interaction is not related to exchanges of identical particles,
but to a single particle being in a superposition of different
position states. The different picture suggested later by
Heisenberg returns to the {\it Austausch} of a proton and a
neutron. A careful discussion of this interpretative issue is
carried out in Refs. \cite{Carson2} and \cite{MillerBook1994} and
references therein.} As we will see in the following, the
introduction of the concept of isospin, which permits
considering the proton and the neutron as different states of
otherwise identical particles, allowed Heisenberg to implement
the idea of exchange between the neutron and the proton without
resorting to the picture of a moving Bose
electron.\footnote{Isospin is just a formal
device that is equivalent to the usual formalism, and then it is not
really crucial in Heisenberg's theory. Majorana
\cite{Majorana1} will not use it in his formulation of exchange
nuclear interactions.} Summing up, the above quote shows
that Heisenberg considered two different exchange mechanisms at the
basis of the interaction that keeps nuclei together, one modeled
after the interaction of the constituent atoms of a molecule, the
other one modeled instead after the electrons in a many-body system. As we
shall see, the second mechanism will underlie the
Majorana theory (Section 3), while the first one will be at the basis
of the subsequent developments based on the Fermi theory (Section
4) and on the meson theory (Section 6), leading to the current
picture of interactions mediated by exchange of virtual particles.
An interesting consideration is in order here
\cite{Carson1,Carson2}. According to the standard interpretation
of the wave function in quantum mechanics, which by 1932 was
already accepted by many physicists, including Heisenberg, the
wave functions of electrons in molecules are to be considered as
stationary states, hence the mental pictures of electrons that
exchange their position or of an electron bouncing back and forth
is approximate at best:\footnote{Such pictures appeared frequently
among the pioneering papers on molecular physics in the late
1920s, but then the mentioned interpretation was not so
established.} there is no net motion of charge in an isolated
molecule. Nevertheless, in 1932 Heisenberg adopted the picture of
the exchange of an electron, and this will lead over the years to
the current representation of fundamental interactions.

Heisenberg assumed that exchange interactions, modeled as we
just described, act between a neutron and a proton (with the exchange
integral denoted by $J(r)$ and assumed to be positive), and that
similar interactions acted between a neutron and a neutron (the
exchange integral in this case denoted by $K(r)$, with the
assumption that $J(r)> K(r)$), but not between a proton and a proton,
where only the Coulomb repulsion $\frac{e^2}{r}$
acts.\footnote{Of course, there is no electron that can be
exchanged in this case. Notice that the charge independence of
nuclear forces had not been discovered yet \cite{PaisIWB}.}
Denoting with $D$ the mass defect of the proton with respect to
the neutron, and finally neglecting all relativistic effects
(allowed by the large mass of all particles), the
complete Hamiltonian for the nucleus proposed by Heisenberg was the
following:
\begin{eqnarray}\label{Hamiltonian1}
H &= & \frac{1}{2M} \sum_k {\bf p}_k^2 - \frac{1}{2} \sum_{k > l} J(r_{kl}) (\rho_k^{\xi} \rho_l^{\xi} + \rho_k^{\eta} \rho_l^{\eta}) - \frac{1}{4} \sum_{k > l} K(r_{kl}) (1 + \rho_k^{\zeta})(1 + \rho_l^{\zeta})   \nonumber \\
& & + \, \frac{1}{4} \sum_{k > l} \frac{e^2}{r_{kl}} (1 - \rho_k^{\zeta})(1 - \rho_l^{\zeta}) - \frac{1}{2} D \sum_k (1 + \rho_k^{\zeta}),
\end{eqnarray}
where $M$ is the proton or neutron mass, $r_{kl} = \left| {\bf
r}_k - {\bf r}_l \right|$ are distances of the order of the
nuclear dimensions and ${\bf p}_k$ is the momentum of the $k$-th
nucleon. This formula involves a fifth quantity for the
characterization of each particle in addition to the usual ones
(i.e. the three position coordinates and the spin $\sigma^z$ along
the $z$-axis), namely a {\it charge} $\rho^{\zeta}$, whose values
are $\rho^{\zeta} = +1$ for a neutron and $\rho^{\zeta} = -1$ for
a proton respectively. The following matrices were introduced
in the space $\xi, \eta, \zeta$:
\begin{eqnarray}\label{matrix}
\rho^{\xi} = \begin{pmatrix}
0 & 1 \\
1 & 0
\end{pmatrix},\
\rho^{\eta} = \begin{pmatrix}
0 & -i \\
i & 0
\end{pmatrix},\
\rho^{\zeta} = \begin{pmatrix}
1 & 0 \\
0 & -1
\end{pmatrix}.
\end{eqnarray}
The fifth quantity $\rho^{\zeta}$ allowed Heisenberg to interpret
the neutron and the proton as the two different states of a
``nucleon", so that the Hamiltonian (\ref{Hamiltonian1}) contains
transition elements which change $\rho^{\zeta} = +1$ into
$\rho^{\zeta} = -1$, i.e. they transfer a {\it charge} between the
neutron and the proton. Later, $\rho^{\zeta}$ would have been
dubbed isotopic spin by Eugene P. Wigner\footnote{Nuclear physicists actually
prefer the term \emph{isobaric} spin, which is in fact more
correct, since two nuclei which differ by the state of one or more
nucleons are termed isobaric, not isotopes.} \cite{Wigner1}.

Then Heisenberg switched to the discussion of the behavior of a
nucleus made of $n$ particles, with $n_1$ neutrons and $n_2$
protons, starting from Eq. (\ref{Hamiltonian1}) and retaining only
the first two terms, which are symmetrical with respect to protons
and neutrons. In this situation the minimum energy corresponds to
a nucleus with $n_1 = n_2$, in agreement with experimental
results. By including the last three terms, the minimum energy
changes, and corresponds to a different situation in which the
ratio between $n_1$ and $n_2$ is higher than one. However, as
Heisenberg noticed, the application to concrete cases needs a
careful discussion of nuclear stability. The solution is immediate
only for the simplest case of hydrogen isotope of weight two,
which is made of one proton and one neutron. The resulting
eigenfunction in the lowest energy state is symmetric in the
spatial and charge coordinates of the two particles, as a
consequence of their spin. This conclusion is a consequence of the
positive sign that Heisenberg assumed for
$J(r)$,\footnote{As in the helium atom or in
molecular physics, the character of exchange interaction depends
on the spin of the state, and the sign of the exchange integral
determines whether the interaction is attractive in a singlet or
in a triplet state.} and is in fact contradicted by experiments,
as we shall discuss later.

Heisenberg carried out only a qualitative discussion of the
general case, starting from the analysis of nuclei containing
neutrons only. On the basis of Eq. (\ref{Hamiltonian1}), a nucleus
with two neutrons is expected to be stable, as well as the $He$
nucleus which is made of two protons and two neutrons. Then, he
addressed the issue of the stability of the nucleus, taken as a
structure containing both neutrons and protons, with the number of
neutrons slightly higher than that of protons. He focused on
intermediate as well as heavy nuclei, and his primary concern was
to find the conditions for stability (against $\beta$- as well as
$\alpha$-decay) and characterize the decay following
instability. In particular he carried out a qualitative study of
the binding energy of the nuclei as a function of $n_1$ and $n_2$.
These topics were deepened in the second paper
\cite{HeisenbergEX2}, where the energy differences between nuclei
with even $n_1$ and $n_2$, odd $n_1$ and $n_2$,  odd $n_1$ or $n_2$, were studied.

In Ref. \cite{HeisenbergEX2}, Heisenberg addressed two further
issues, the scattering of $\gamma$-rays by atomic nuclei and the
properties of the neutron, which will be the topic of the third
paper of the series, published in 1933 \cite{HeisenbergEX3}. With
respect to the scattering of $\gamma$-rays, Heisenberg's
explanation made crucial use of the assumption of the
neutron as a bound state of a proton and an electron. The
composite structure of the neutron implies that not only the
motion of the protons and neutrons changes due to the effect of
the external radiation, but also that the nuclear electrons inside
neutrons get excited. Though wrong, these hypotheses lead him
to a satisfactory explanation of the anomalous scattering of
$\gamma$-rays on nuclei (known as the Meitner-H\"{u}pfeld effect
\cite{Tarrant,MeitnerH,Chao}). The second mechanism is more
effective thanks to the small mass of the electron and gives a
significant contribution to the total scattering, which could be
in agreement with experimental data.\footnote{It would
later be established that the anomalously large cross section is
due to the creation of electron-positron pair from the $\gamma$
photon in the intense electric field close to the nucleus
\cite{BrownMoyer}.} Concerning the properties of the neutron,
Heisenberg made a comparison between the two hypotheses on the
nature of this particle (elementary particle without structure
versus composite system of a proton and an electron). While the
second hypothesis seemed to work well both in the case of
$\beta$-decay and in explaining the nature of exchange interaction
within the nuclei, it brought about several conceptual difficulties,
which led him again to question the validity of quantum mechanics
at the nuclear scale. In particular, the uncertainty principle
gives for the mass defect of the neutron an energy of the order of
$137 \ mc^2$, while the observed mass defect is about a hundred
times smaller and has the opposite sign. Furthermore, quite
different values of the binding energy of the nuclear electron are
obtained depending on the experiments devised to measure it.

Heisenberg's third paper on nuclear forces \cite{HeisenbergEX3}
further extends the ideas introduced in the previous papers
\cite{HeisenbergEX1, HeisenbergEX2} starting from a generalized
Hamiltonian, obtained by augmenting Eq. (\ref{Hamiltonian1}) with
the term $+ \frac{1}{2} \sum_{k > l} L(r_{kl}) (1 - \rho_k^{\zeta}
\rho_l^{\zeta})$, which adds ordinary {\it
static} interactions between protons and neutrons to the exchange forces. Then, an
approximate solution was found by applying the Thomas-Fermi method to the case of
many-particle nuclei. In this method, the
nucleus is modeled as a gas of free particles which obey Fermi
statistics, bound together by forces whose potential energy
is:
\begin{eqnarray}\label{TFpotential1}
U(r) = -2 \frac{n_1 n_2}{n(n-1)} \left[ J(r) - L(r) \right] - \frac{n_1(n_1-1)}{n(n-1)} K(r) + \frac{n_2(n_2-1)}{n(n+1)} \frac{e^2}{r},
\end{eqnarray}
while the total energy in the lowest state is:
\begin{eqnarray}\label{TFenergy1}
E =  \frac{h^2}{M} \frac{4\pi}{5} \left( \frac{3}{8\pi} \right)^{\frac{5}{3}}\int \!\! \rho({\bf r})^{\frac{5}{3}} d \tau
+  \frac{1}{2} \int \!\!\!\! \int \!\! \rho({\bf r})\rho({\bf r}^{'}) U(|{\bf r}-{\bf r}^{'}|) d \tau d \tau^{'} -  n_1 D.
\end{eqnarray}
Here the density distribution $\rho({\bf r})$ can be found by
minimizing $E$ under the constraint $\int \!\!\rho({\bf r}) d\tau =
n$. The drawback of this expression is that, for values of $|{\bf
r}-{\bf r}^{'}|$ such that the effective potential $U(|{\bf
r}-{\bf r}^{'}|)$ suddenly increases, the double integral may
develop a divergence. This means that, in order to have
saturation, the effective potential $U(|{\bf r}-{\bf r}^{'}|)$ has
to be strongly repulsive at short distances. Technically,
Heisenberg obtained the result by putting $U(|{\bf r}-{\bf
r}^{'}|)$ equal to zero for values of $|{\bf r}-{\bf r}^{'}|$
smaller than the minimum separation between two particles.

Summing up, it is clear that despite being a milestone in
the quantum theory of nuclear structure, Heisenberg's work on
nuclear forces suffers from a number of drawbacks. The analysis of
these drawbacks was the starting point of Majorana's work.

\section{Majorana's contribution on nuclear forces}

Before talking about Majorana's celebrated contribution
\cite{Majorana1,Recami1,Esposito1,Brink1}, we focus on the
scientific environment in which his work was conceived and
developed. The circumstances which brought Majorana to deal with
the issues of the nature of neutron and the exchange forces can be
found in a detailed account by his colleagues working in Rome
under the leadership of Fermi, i.e., mainly Edoardo Amaldi
\cite{Recami1,Amaldi1,Amaldi2,Recami2}. The story went on as
follows. 

The beginning of 1932 was marked by the experiments by
Fr\'ed\'eric Joliot and Ir\`ene Curie \cite{JoliotCurie1}, which
Majorana soon understood as the very discovery of the {\it neutral
proton} (e.g. the neutron) even if the Joliot-Curies did not
realize it. Hence, Majorana succeeded in providing an explanation
of the structure and the stability of atomic nuclei only in terms
of neutrons and protons, and this probably happened, as vividly
remembered by Amaldi \cite{Amaldi1,Amaldi2}, before both the
announcement of the discovery of the neutron by Chadwick
\cite{Chad1932a} and the work by Dmitrij D. Ivanenko
\cite{Iwanenko}:

\begin{quote}
Soon after Chadwick's discovery, various authors understood that
the neutron must be one of the components of the nucleus and began to
propose various models which included alpha particles, protons, electrons and
neutrons. The first to publish the idea that the nucleus consists solely of
protons and neutrons was probably Iwanenko [...] What is
certain is that before Easter of that year [Ettore Majorana]
tried to work out a theory on light nuclei, assuming that they consisted solely of protons
and neutrons (or ``neutral protons'' as he then said) and that the former interacted with the
latter through exchange forces. He also reached the conclusion that these exchange forces must act only on the space coordinates
(and not on the spin) if one wanted the alpha particle, and not the deuteron, to be the system saturated with respect to binding energy
(\cite{Amaldi3}, p. 31).
\end{quote}
Majorana gave a brief account of his theory to Fermi and his group
but he did not proceed to publish his results, despite Fermi's
advice. Heisenberg's first work on exchange forces appeared in the
\emph{Zeitschrift f\"{u}r Physik} in July 1932, but it was only
later that Majorana, strongly encouraged by Fermi, applied for a
grant of the Italian National Research Council (CNR) in order to
spend a semester abroad, first in Leipzig and then in Copenhagen,
to meet Heisenberg. During his research visit in Leipzig, started
in January 1933, he was convinced by Heisenberg to publish his
results on nuclear theory in a paper, which appeared in the
\emph{Zeitschrift f\"{u}r Physik} in March 1933. 

Majorana's contribution starts with an enlightening critical
analysis of Heisenberg's theory:

\begin{quote}
In order to find a suitable interaction between the components of
the nuclei Heisenberg was guided by an obvious analogy. He treats
the neutron as a combination of a proton and an electron, i.e.
like a hydrogen atom bound by a process not fully understood by
present theories, in such a way that it changes its statistical
properties and its spin. He further assumes that there are
exchange forces between protons and neutrons similar to those
responsible for the molecular binding of $H$ and $H^+$. [...] One
may doubt the validity of this analogy as the theory does not
explain the inner structure of the neutron, and the interaction
between neutron and proton seems rather big compared with the
mass-defect of the neutron as determined by Chadwick. We think,
therefore, that it may be quite interesting to find a Hamiltonian
very similar to Heisenberg's which represents in the simplest way
the most general and most obvious properties of the nucleus. [...]
If we assume that nuclei consist of protons and neutrons we have
to formulate the simplest law of interaction between them which
will lead, if the electrostatic repulsion is negligible, to a
constant density for nuclear matter (\cite{SIF}, pp. 186-187).
\end{quote}
Here Majorana clearly states his program, pointing out his search
for the {\it simplest law} which could describe the neutron-proton
interaction and provide the correct observed properties of nuclei,
mainly the so called saturation. This may be obtained in whole
analogy to the structure of solid and liquid matter by taking an
attractive force for large distances and a short range strong
repulsive force in order to take into account impenetrability of
particles. But this solution is considered by Majorana
``aesthetically unsatisfactory'', so that he concludes:

\begin{quote}
We shall, therefore, try to find another solution and introduce as
few arbitrary elements as possible. The main problem is this: How
can we obtain a density independent of the nuclear mass without
obstructing the free movement of the particles by an artificial
impenetrability? We must try to find an interaction whose average
energy per particle never exceeds a certain limit however great
the density. This might occur through a sort of saturation
phenomenon more or less analogous to valence saturation
(\cite{SIF}, p. 188).
\end{quote}
Accordingly, he devised an exchange interaction which had an
opposite sign with respect to that postulated by Heisenberg and,
in particular, which exchanged only the position coordinates of
proton and neutron (while Heisenberg's interaction exchanged both
position and spin coordinates). Following Dirac
\cite{Dirac1930}, he writes down this interaction as:
\begin{eqnarray}\label{Majorana1}
\left(Q^{'}, q^{'} \left| J \right| Q^{''}, q^{''} \right) =  - \delta \left(q^{'} - Q^{''}  \right) \delta \left(q^{''} - Q^{'}  \right) J(r).
\end{eqnarray}
Here $r = \left| q^{'} - Q^{'} \right|$, where $Q$ and $q$ are the
position coordinates of a neutron and a proton. Regarding $J(r)$,
it is taken positive and only at the end of the paper Majorana
hints to its functional form, to be chosen in order to reproduce
experimental data. At variance with Heisenberg's exchange force,
which saturates in the deuteron, Majorana's force, Eq.
(\ref{Majorana1}), allows each neutron (proton) to bind to two
protons (neutrons), so that saturation takes place in the
$\alpha$-particle:

\begin{quote}
Thus we find that both neutrons act on each proton in the
$\alpha$-particle instead of only one and vice versa, since we
assume a symmetrical function in the position coordinates of all
protons and neutrons (which is true only if we neglect the Coulomb
energy of the protons). In the $\alpha$-particle all four
particles are in the same state so that it is a closed shell. If
we proceed from an $\alpha$-particle to heavier nuclei we can have
no more particles in the same state because of the Pauli
principle. Also, the exchange energy is usually large only if a
proton and a neutron are in the same state and we may expect,
which agrees with experiments, that in heavy nuclei the mass
defect per particle is not noticeably bigger than in the
$\alpha$-particle (\cite{SIF}, pp. 189).
\end{quote}
These stability properties of Majorana's exchange force gained a
wide consensus within the scientific community, starting from the
endorsement which he received by Heisenberg during his speech at
the Solvay Conference, held in October 1933
\cite{HeisenbergSolvay1933}. There, Heisenberg expressed Majorana's
exchange force (\ref{Majorana1}) as:
\begin{eqnarray}\label{Majorana2}
\frac{1}{4} J(r_{kl}) (\rho_k^{\xi} \rho_l^{\xi} + \rho_k^{\eta} \rho_l^{\eta}) (1 + \sigma_k \sigma_l),
\end{eqnarray}
or, without introducing isotopic spin, as:
\begin{eqnarray}\label{Majorana3}
- J(r_{kl}) P_{kl},
\end{eqnarray}
$P_{kl}$ being an operator giving the permutation of the
spatial coordinates $r_k$ and $r_l$. Then he compared his own
expressions for exchange forces,
\begin{eqnarray}\label{Heis1a}
\frac{1}{2} J(r_{kl}) (\rho_k^{\xi} \rho_l^{\xi} + \rho_k^{\eta} \rho_l^{\eta}),
\end{eqnarray}
or
\begin{eqnarray}\label{Heis1b}
- J(r_{kl}) P_{kl}^{'},
\end{eqnarray}
($P_{kl}^{'}$ being an operator permuting spatial as well as
spin coordinates $r_k$, $\sigma_k$ and $r_l$, $\sigma_l$) with the
ones by Majorana, Eqs. (\ref{Majorana2})-(\ref{Majorana3}), and
soon recognized the validity of Majorana's approach.

After pointing out the qualifying points of his approach, Majorana
switched to the explicit proof of the density saturation of the
nuclear components, which had been found experimentally. The
calculation is carried out by applying the Thomas-Fermi model to
nuclear matter, as had also been done by Heisenberg in his third
paper on nuclear forces \cite{HeisenbergEX3}. As a starting point,
the eigenfunction $\psi$ of the nucleus, built up of $n_1$
neutrons and $n_2$ protons, is written as a product of two
functions, $\psi_N$ and $\psi_P$, dependent on the coordinates
(position and spin) of the $n_1$ neutrons and $n_2$ protons,
respectively. In turn, $\psi_N$ and $\psi_P$ can be written as
Slater's determinants, constructed starting from orthogonal single
particle eigenfunctions. Then the calculation proceeds by taking
the average value of the total energy $W$ over the chosen wave
function $\psi$ and then looking for its minimum. $W$ is taken as
the sum of the kinetic energy, the electrostatic energy of the
proton and the exchange energy. By introducing the Dirac density
matrices \cite{Dirac1930} $\rho_N$ and $\rho_P$ for
neutrons and protons, respectively, and then neglecting the Coulomb
repulsion among protons, while fixing the ratio $\frac{n_1}{n_2}$,
the potential energy per particle reduces to the exchange energy
contribution and can be written as a function $a(\mu)$ of the
total nuclear density:
\begin{eqnarray}\label{TMnuclearDensity1}
\mu = \frac{8 \pi}{3 h^3} \left(P_N^3 + P_P^3\right),
\end{eqnarray}
where, as usual within the Thomas-Fermi framework, $P_N$ and $P_P$
are the Fermi momenta of neutrons and protons. The behavior of
$a(\mu)$ as a function of density is studied starting from the two
limiting cases $\mu = 0$ and $\mu \rightarrow \infty$. It vanishes
in the first case, while approaching the constant value $- \frac{2
n_2}{n_1 + n_2} J(0)$ in the second case. For an
intermediate density, the following expression for $a(\mu)$ is
derived:
\begin{eqnarray}\label{exchangeEnergyTM}
a(\mu) = \frac{1}{\mu(q)} \int \!\!\!\!\int \! \frac{\rho_N (p,q) \rho_P (p^{'},q)}{h^6} G(p,p^{'}) dp dp^{'},
\end{eqnarray}
where
\begin{eqnarray}\label{Gfunction}
G(p,p^{'}) = \int \!\!e^{-\frac{2\pi i}{h} (p - p^{'},v)} J|v| dv,
\end{eqnarray}
while the kinetic energy per particle is $\kappa \mu^{2/3}$. Then
Majorana was able to show that there exists a minimum of the total
energy $a+t$ per particle for a particular value $\mu_0$ of the
density, which depends only on the ratio $\frac{n_1}{n_2}$ and
corresponds to a stable nucleus.

Finally he focused on the search for the best analytic
form of the function $J(r)$, which could be in agreement with
experimental data. His first proposal is $J(r) = \lambda
\frac{e^2}{r}$, dependent on one arbitrary constant and showing a
divergence at $r = 0$, but it is soon discarded:

\begin{quote}
For great distances, however, it must be modified as it gives an
infinite cross section for the collision between protons and
neutrons. Also, it seems to provide too small a ratio for the mass
defects of the $\alpha$-particle and the hydrogen isotope. Thus,
we have to use an expression with at least two constants, e.g. an
exponential function, [...] (\cite{SIF}, p. 193).
\end{quote}
Thus he proposed the two constants function $J(r) = A e^{-\beta
r}$, but apparently he didn't pursued the calculations further. In
his words:

\begin{quote}
We shall not follow this up since it has been shown that the first
statistical approximation can lead to considerable errors however
large the number of particles (\cite{SIF}, p. 193).
\end{quote}
While Majorana did not pursue this matter in the published
work, unpublished research notes \cite{Esposito2} by the same
author show that he did in fact carry out calculations and
provided numerical estimates for both proposals, also trying to
further generalize his work.

Let us notice that the same two constants function for $J(r)$ was
later taken by Heisenberg \cite{HeisenbergSolvay1933} to carry out
Thomas-Fermi calculations for a nucleus built of a large number of
constituents, in whole analogy with Majorana. A similar two
constants expression, $J(r) = - g^2 \frac{e^{- \lambda r}}{r}$,
would have been introduced by Yukawa in 1935, $g$ being the
coupling constant for the interaction between neutron, proton and
the so called $U$-particle
\cite{Yukawa1935,Brown1,Brown2,Darrigol1}.

But neither Heisenberg nor Majorana addressed the problem of
energy non-conservation in $\beta$-decay. Furthermore, even if
Majorana's formulation of exchange forces gained a wide consensus,
it was Heisenberg's isotopic spin framework that became widespread
in nuclear physics, and it was used in particular in Fermi's
theory of $\beta$-decay, to which we now turn.

\section{The  solution to the $\beta$-decay puzzle: Fermi's theory}

The $\beta$-decay puzzle found a solution in 1934, when Fermi
published a theory where the process was modeled in a way
strongly relying on the analogy with the emission of a photon by a
decaying atom \cite{Fermi1,Fermi2,Fermi3,Fermi4}. Fermi's quantum
theory of $\beta$-decay put together Pauli's neutrino hypothesis,
Heisenberg's isotopic spin framework and second quantization
\cite{JordanWigner}. It provided expressions for the
lifetime and the shape of the $\beta$-ray emission spectra to be
compared with experimental data. As Tomonaga pointed out in his
book ``The story of spin'':

\begin{quote}
[...] even Heisenberg apparently was not free from Bohr's
influence, and he dared not touch the problem of $\beta$-decay,
believing that quantum mechanics was not applicable. It was none
other than Fermi who rejected Bohr's idea and incorporated
$\beta$-decay into the framework of quantum mechanics utilizing
Pauli's neutrino (\cite{TomonagaSpin}, p. 176).
\end{quote}
Fermi's starting point was, besides neutrino's hypothesis, the
assumption that nuclear constituents are only neutrons and protons
(heavy particles). The qualifying point of his strategy is the
analogy with the electromagnetic case, as clearly stated at the
beginning of his work:

\begin{quote}
[...] in order to understand that $\beta$ emission is possible, we
want to try to construct a theory of the emission of lightweight
particles from the nucleus in analogy with the theory of emission
of light quanta from an excited atom by the usual radiation
process. In radiation theory, the total number of light quanta is
not constant. Light quanta are created when they are emitted from
an atom, and are annihilated when they are absorbed
(\cite{Fermi4}, p. 1151).
\end{quote}
These considerations on the electromagnetic case led Fermi to
assume that also in $\beta$-decay the total number of electrons is
not constant, and that the same holds for neutrinos. This implies
the possibility both for electrons and neutrinos to be created and
annihilated, thus getting rid at once of all the problems
associated with electrons being confined in the nucleus.
Furthermore neutron and proton are considered, as done by
Heisenberg, as two internal quantum states of the nucleon, so that
a isospin coordinate $\rho$ is introduced, such that $\rho = 1$
for the neutron and $\rho = -1$ for the proton. Finally, the
Hamiltonian of the system including heavy and light particles can
be constructed in order to implement the following processes, while
satisfying charge conservation: a transition from a neutron to a
proton accompanied by the creation of an electron and a neutrino
and the inverse transition (proton to neutron) together with the
annihilation of an electron and a neutrino.

As a result of the above guidelines, the following Hamiltonian is
introduced, as a sum of the energy of the heavy particles
$H_{hp}$, that of the lightweight particles $H_{lp}$, and finally the
interaction energy $H_{int}$ between the heavy and lightweight
particles:
\begin{eqnarray}\label{FermiHam1}
H_F = H_{hp} + H_{lp} + H_{int}.
\end{eqnarray}
Here the first term is written for a single heavy particle as:
\begin{eqnarray}\label{FermiHam2}
H_{hp} = \frac{1}{2} (1 + \rho) N + \frac{1}{2} (1 - \rho) P,
\end{eqnarray}
where $\rho$ is the isospin variable, $N$ and $P$ are the energy
operators of the neutron and the proton, respectively. For the
second term one has:
\begin{eqnarray}\label{FermiHam3}
H_{lp} = \sum_s H_s N_s + \sum_{\sigma} K_{\sigma} M_{\sigma},
\end{eqnarray}
where $H_1, H_2, \dots , H_s, \ldots$ and $K_1, K_2,
\ldots, K_{\sigma}, \ldots$ are the energies of electrons
and neutrinos, respectively, while  $N_1, N_2, \ldots, N_s,
\ldots$ and $M_1, M_2, \ldots, M_{\sigma}, \ldots$ are
the occupation numbers of the individual quantum states $\psi_1,
\psi_2, \ldots, \psi_s, \ldots$ of the electrons and
$\phi_1, \phi_2, \ldots, \phi_{\sigma}, \ldots$ of the
neutrinos. Annihilation and creation operators $a_s$ and $a_s^{*}$
for electrons have been introduced such that $\psi = \sum_s \psi_s
a_s$ and $\psi^{*}= \sum_s \psi_s^{*} a_s^{*}$. Similar
operators, $b_{\sigma}$ and $b_{\sigma}^{*}$, have been introduced
for neutrinos such that $\phi = \sum_{\sigma} \phi_{\sigma}
b_{\sigma}$ and $\phi^{*}= \sum_{\sigma} \phi_{\sigma}^{*}
b_{\sigma}^{*}$.

Finally, for the interaction term the following simple choice has
been adopted by neglecting relativistic corrections and spin
interaction:
\begin{eqnarray}\label{FermiInteraction}
H_{int} = g \left[ Q \psi (x) \phi (x) + Q^{*} \psi^{*} (x) \phi^{*} (x) \right],
\end{eqnarray}
where $g$ is a dimensional constant, $x$ are the coordinates of
heavy particles, such that the field operators $\psi$, $\phi$,
$\psi^{*}$, $\phi^{*}$ have to be evaluated at the position of
heavy particles, while $Q = \frac{1}{2} (\rho^{\xi} -
i\rho^{\eta}) = \begin{pmatrix} 0 & 1 \\ 0 & 0 \end{pmatrix}$ and
$Q^{*} = \frac{1}{2} (\rho^{\xi} + i\rho^{\eta}) = \begin{pmatrix}
0 & 0 \\ 1 & 0 \end{pmatrix}$ are operators implementing the
transition from a proton to a neutron,as well as that from a neutron to a
proton, acting on the functions of the isospin
variable $\rho$.

The interaction term was also generalized by Fermi to the case in
which at least for the lightweight particles a relativistic
treatment could be carried out. This required the introduction of
Dirac 4-spinors $\psi$ and $\phi$ and their Hermitian conjugates
$\tilde{\psi}$ and $\tilde{\phi}$ for electrons and neutrinos,
respectively, but the velocities of heavy particles are usually
small if compared with the speed of light $c$, so that only the
scalar potential term is retained. The resulting interaction term
is:
\begin{eqnarray}\label{FermiInteraction1}
H_{int} = g \left[ Q \tilde{\psi}^{*} \delta \phi + Q^{*} \tilde{\psi} \delta \phi^{*} \right],
\end{eqnarray}
where $\delta$ is a matrix defined as $$\begin{pmatrix} 0 & -1 & 0
& 0 \\ 1 & 0 & 0 & 0 \\ 0 & 0 & 0 & 1 \\ 0 & 0 & -1 & 0
\end{pmatrix}.$$

With the above Hamiltonian the theory of $\beta$-decay was fully
developed in whole analogy with quantum radiation theory, by
carrying out perturbative calculations with the interaction term
$H_{int}$ taken as perturbation. A comparison with available
experimental data was performed, unveiling a good agreement with
theory, as recognized by Fermi himself:

\begin{quote}
To summarize, one can say that this comparison of theory and
experiment gives as good an agreement as one could expect
(\cite{Fermi4}, p. 1159).
\end{quote}
Fermi provided also an estimate of the coupling constant $g$,
whose order of magnitude was found to be $4 \cdot 10^{-50}$ erg
cm$^{3}$.

The success of Fermi's theory of $\beta$-decay triggered
subsequent work towards the application of his ideas to the
mechanism of exchange force, in which an electron-neutrino pair
could play the role of the Bose electron. This picture was
implemented by Heisenberg, who mentioned his results in a letter
written to Pauli on 18 January 1934 \cite{HeiPauliLetter1934}. By
carrying out a second order perturbative calculation, he found that the
Fermi interaction would give rise to an exchange energy $J(r)
\simeq \frac{const}{r^5}$, which is very small for $r \le \frac{\hbar}{Mc}$
and does not match the order of magnitude of the
interaction of neutron and proton observed in experiments.
Subsequent independent work by Igor Tamm \cite{Tamm1} and Ivanenko
\cite{Iwanenko1} confirmed Heisenberg's negative results,
demanding for new ideas. The puzzle would have been solved by
Yukawa one year later, as we will see below.

\section{Brief interlude: theoretical physics and quantum mechanics in Japan from the late 1920s to the 1930s}

Here we briefly summarize the state of art of
theoretical physics and, in particular, of quantum mechanics in
Japan, starting from the late 1920s (for further details, see
e.g. \cite{Ito1,Ito2} and references therein). The focus is also
on the scientific exchanges with European physicists, which
contributed to shape the work and the career of Yukawa
\cite{Tabibito} and Tomonaga \cite{TomonagaSpin}.

After the forced opening of the country to international trade in
1854 and the re-establishment of the emperor in 1868, a number of
Japanese scientists was sent abroad to study science, engineering
and medicine, with the aim to bring back in Japan what they had
learnt. In the same period many universities were founded in
T\={o}ky\={o}, Ky\={o}to and other places. Nagaoka Hantar\={o} was
the first physicist to go to Europe, in 1893. Back in Japan,
in 1904 he proposed a ``Saturnian'' model of the atom, built of a
positive nucleus surrounded by a ring of negative electrons, which is historically the first atomic model containing a nucleus. In
the meantime, research laboratories were being established
throughout the country, including the Institute of
Chemical and Physical Research (RIKEN) laid down in T\={o}ky\={o}
in 1917 and funded also by government. Although its mission was to
support industries with applicative and technological studies,
basic research was carried out there too. In 1921 Nishina
Yoshi\={o}, a student of Nagaoka, went abroad, in Germany, England
and mainly at the Niels Bohr Institute in Copenhagen, where he spent
six years, thus witnessing the unfolding of the quantum
revolutions and greatly contributing himself.\footnote{His most
well-known result is the celebrated Klein-Nishina formula
\cite{KNformula}. See Tomonaga's recollections (in
\cite{TomonagaSpin}, pp. 220-221) for an interesting account of
Nishina's experience in deriving his results in collaboration with
Oskar Klein.} Back in Japan in 1928, he brought in his country the
Copenhagen spirit, which meant a new style in doing research, and
the ideas of new quantum mechanics. Theoretical
physics began to wide spread in Japan just thanks to Nishina. It is also worth to mention
the famous Einstein visit in 1922 in Japan \cite{EinsteinV}, where
he met Nagaoka, Tamaki Kajuro and Ishiwara Jun, who would
have published a book on relativity and contributed to promote
Einstein's theories in their country. All these scientists were in charge
at Ky\={o}to University when Yukawa and Tomonaga would have been
undergraduate students. In 1931, Nishina established his
laboratory at RIKEN, where he trained many young physicists,
including Yukawa and Tomonaga, and where experimental work on cosmic rays
and nuclear physics was performed too. 

\section{Yukawa and the meson theory of nuclear forces}

In this Section we introduce Yukawa's contribution to the theory
of nuclear forces, but, as a preliminary step, a short biographical
sketch, including his early interactions with Tomonaga, is
given \cite{Tabibito,Brown2,Spradley,Sato1,KobayasiM}.

\subsection{Short biographical notes}

Born in T\={o}ky\={o} on January 23, 1907, Ogawa Hideki was the
third son of a staff member of the governmental Geological Survey
Bureau. In 1908 his father was appointed professor of Geography at
the Faculty of Arts of Ky\={o}to University, so Hideki and his
family moved to Ky\={o}to. He began to study the Chinese classics
with his grandfather, later discovering (when adolescent) the
writings of Taoists sages. He was strongly attracted by a sort of
dialectical materialism, very different from Marx' and Engels'
ideas: with placing nature at the center of the
universe, indeed, Taoism strongly influenced the young Yukawa's thinking. In 1923
he was enrolled into the Third High School while in 1926 he began
his physics studies in Ky\={o}to Imperial University. He had Tomonaga as
classmate since high school, and both went to study
physics in Ky\={o}to. Hideki began studying quantum mechanics also by
checking the academic journals available in the library of the
physics department, thus reading original research papers
(\cite{TomonagaSpin}, p. 222). After the graduation in 1929, both
Yukawa and Tomonaga continued to study quantum mechanics by themselves 
as unpaid research assistants in Ky\={o}to, while Nishina
had just returned to Japan after his research visit in Europe.
Besides lecturing on the new physics, Nishina was the main
organizer of a cycle of lectures held in Ky\={o}to and
T\={o}ky\={o} by Heisenberg and Dirac
\cite{Konagaya,Ito1,Ito2,Kim1}. The aim was to engage young
physics students and stimulate them to pursue research on new and
cutting edge topics.\footnote{Some detail about these lectures is
given in Tomonaga's recollections (\cite{TomonagaSpin}, p. 222),
where it is explicitly stated that Heisenberg lectured in
T\={o}ky\={o}, both at University and at RIKEN, from September 2
to 9 (1929).The titles of his lectures are as follows: 1) ``Theory of Ferromagnetism''; 2)
``Theory of Conduction'' (Bloch's theory of electric conductions);
3) ``Retarded Potential in the Quantum Theory'' (Heisenberg-Pauli
theory); ``The Indeterminacy Relations and the Physical Principles
of the Quantum Theory''. Dirac's lectures at University of
T\={o}ky\={o} were: 1) ``The Basis of Statistical Quantum
Mechanics'' (the density matrix); 2) ``Quantum Mechanics of
Many-Electron Systems'' (i.e. the representation of permutation
operator of electron coordinates in terms of spin variables and
applications); 3) ``Relativistic Theory of the Electron'' (i.e.
the Dirac equation); while the last one, 4) `The Principle of
Superposition and the Two-Dimensional Harmonic Oscillator'' , was
given at RIKEN. } Without any doubt, these events played a crucial
role in shaping the future scientific career of both Yukawa and
Tomonaga, who attended those lectures. As Yukawa recalled in his autobiography
\cite{Tabibito}:

\begin{quote}
During the two years following my graduation, Bunsaku Araktsu,
Yoshikatsu Sugiura, and Yoshio Nishina lectured on quantum
mechanics, presenting different points of view. All of them had
studied the new physics in Europe; Professor Nishina had the
greatest influence on us. At that time, the phrase ``Copenhagen
spirit'' was frequently heard in the physics world, referring to
the Institute of Theoretical Physics at Copenhagen University,
with Niels Bohr as its head. The best theoretical physicists came
from all over the world to learn from Bohr, including some
Japanese scientists. Yoshio Nishina had a particularly long stay
in Copenhagen. His lectures were not only explanations of quantum
physics, for he carried with him the spirit of Copenhagen, the
spirit of that leading group of theoretical physicists with Niels
Bohr as its center. If I asked to describe the spirit of
Copenhagen, I would not be able to do so in a few words. However,
it is certain that it had much in common with the spirit of
generosity. Having been liberally educated, I was especially
attracted by that, but I was also attracted by Professor Nishina
himself (\cite{Tabibito}, pp. 292-293).
\end{quote}
Hideki started to work as an unpaid research associate in
theoretical physics in professor Tamaki Kajuro's lab, and in the
following years he studied and translated Heisenberg's and Dirac's
papers \cite{Gaudenzi}. In 1932 he married Yukawa Sumiko and was
adopted by the family of his wife, taking also the new family
name. Thus he moved to \={O}saka, while appointed lecturer at
Faculty of Science in Ky\={o}to, where he started teaching quantum
mechanics and carrying out research in nuclear and cosmic ray
physics. Among his students there were Sakata Sh\={o}ichi,
Kobayashi Minoru and Taketani Mitsuo, who later would collaborate
with him and contribute to the development of the new theory of
nuclear forces \cite{Tabibito}. In April 1934 he moved to
\={O}saka University, where he built up and led his new research
group. Those years were very productive, as we will see in the
following sections. Back to Ky\={o}to Imperial University since
1938, he became a member of the National Research Council and
continued carrying out research on meson theory and nuclear
physics while pursuing the new field of elementary particle
physics. In 1939 Yukawa made his first travel abroad, in Europe,
on the occasion of an invitation to Solvay Conference. In the same
period Tomonaga, driven by Nishina's advice, was carrying out
research in Leipzig in Heisenberg's group.\footnote{See Ref.
\cite{DarrigolTom} for a scientific biography of Tomonaga, where a
detailed account of the Leipzig years is given.} In 1948 Yukawa
joined the Institute for Advanced Study in Princeton as a visiting
professor, upon invitation by J. Robert Oppenheimer, and his research 
mainly focused on non-local field theory and extended models
of elementary particles. Appointed professor at Columbia
University in 1949, in the same year he was awarded the Nobel
Prize in physics \cite{YukawaNobel} and elected to the Science
Council of Japan. In 1953 Yukawa returned in Japan and became the
first director of the newly established Research Institute for
Fundamental Physics in Ky\={o}to, a position which he held until
his retirement in 1970.

\subsection{Preliminary work on nuclear forces}

A turning point in Yukawa's scientific career came in 1932, as he
recalled in his autobiography \cite{Tabibito} when talking about
the scientific events of that year:

\begin{quote}
The year 1932 was even more turbulent for physics than it was for
my personal life. Events, each of which, taken alone, could be
called revolutionary, occurred three in a row: first, there was
the discovery of the neutron; second, the discovery of the
positron; third, the atomic nucleus was broken up by artificial
means, namely by the use of the particle accelerators. The
discipline that is now called nuclear physics was until then a
minor branch of study, but because of those three events, it
suddenly became the mainstream (\cite{Tabibito}, pp. 310-311).
\end{quote}
According to Yukawa, the discovery of neutron was a very
significant event for theoretical physics, because it brought into
play the idea that the nucleus is built of protons and neutrons.
Heisenberg's papers on nuclear structure
\cite{HeisenbergEX1,HeisenbergEX2,HeisenbergEX3} deeply impressed
the young Yukawa, who wrote an introduction to the Japanese
translation of these papers to be published in the Proceedings of
the Physico-Mathematical Society of Japan \cite{Tabibito,Brown2}.
This event marked the beginning of his struggle towards the
building of his new theory of nuclear forces. In Yukawa's own
words:

\begin{quote}
The problem that I focused on was that the nature of the forces
that act upon the neutrons and protons making up the nucleus --
that is, the nature of the nuclear forces. By confronting this
difficult problem, I committed myself to long days of suffering.
[...] The fact that I was suffering, however, was very satisfying
to me; I felt like a traveler carrying a heavy burden and
struggling up a slope (\cite{Tabibito}, p. 312).
\end{quote}
His starting point was the consideration that, after the
introduction of quantum mechanics, the only primary (or
fundamental) forces in nature were identified with gravity and
electromagnetism. As a main and common feature, both forces were
expressed in terms of fields. Thus, an answer had to be provided to
the question whether nuclear forces were additional primary ones
or not. In principle, in fact, the weakness of both gravity and
electromagnetic force prevents them to be the source of nuclear
force. Furthermore, electromagnetism could not account for the
attraction of protons within the nucleus as well as for the
interactions of neutrons, which are electrically neutral. Thus the
consideration that strange things happen when thinking to the
nuclear force as a secondary one, to be derived from
electromagnetism, led Yukawa to view nuclear forces as new
fundamental ones and, as such, described by means of a suitable
field. Quantum mechanics then played a crucial role, as recalled
by Yukawa in his autobiography, in driving him to conceive the
idea of meson. But, despite having conceived the general idea
early on, the actual implementation of his program would turn out
to be very hard, and lasted two years:

\begin{quote}
I had this idea of a nuclear force field very early. Looked at
from the quantum mechanical viewpoint, a field of force, almost by
necessity, implies that there is a particle accompanying that
field. [...] I had to take a wrong path first, before I could
arrive at my destination. [...] In retrospect, I was very close to
my destination in 1932. Had I pursued the notion of the field of
nuclear force, and applied quantum mechanical reasoning, I should
have been able to come up with the idea of the meson. Instead, I
spent the next two years searching in the dark (\cite{Tabibito},
pp. 320-321).
\end{quote}
Yukawa started his investigations by following Heisenberg's idea
of the nuclear force as an exchange of an electron between protons
and neutrons \cite{HeisenbergEX1} and presented some preliminary
results at the meeting of the Physico-Mathematical Society of
Japan held in April 1933 at the T\={o}hoku University in Sendai.
He gave a talk entitled ``A Comment on the Problem of Electrons in
the Nucleus'', but he did not publish the corresponding manuscript
in a journal. This and his subsequent work in 1933-34, most of
which is unpublished, were an important preliminary step towards
the research he would have carried out in the following years, and
as such are of great historical interest in order to reconstruct
his path to meson theory.\footnote{Historical materials related to
Yukawa, ranging from calculation notes to manuscripts' drafts and
laboratory records, are collected in the Yukawa Hall Archival
Library (YHAL) at Yukawa Institute for Theoretical Physics (YITP),
Ky\={o}to University \cite{Archive1}. Some documents are also
available in the Archive of historical materials, \={O}saka
University Yukawa Memorial and available via web
\cite{YukawaArchive}. A very useful English translation of some
key archival documents related to Yukawa's preliminary works
on nuclear forces and relativistic quantum field theory has
been given by Kawabe Rokuo in Ref. \cite{Kawabe1}. } In Sendai, Yukawa also met Yagi Hidetsugu, head of the Physics
Department of \={O}saka University, and this circumstance led him to
move to \={O}saka to join the nuclear research group, which had a Cockroft-Walton accelerator. He recalled that ``it was very
stimulating to know that experiments would take place
simultaneously with the theoretical studies'' (\cite{Tabibito}, p.
326).

The 1933 talk contains Yukawa's first idea of an exchange force
due to an electron shuttling between the neutron and the proton, along with
a careful analysis of the many difficulties he met. In
the abstract\footnote{The English translation of the
abstract of the 1933 talk can be found in Ref. \cite{Kawabe1}, p.
247} we can find his program:

\begin{quote}
We can consider that, by analogy with radiation (in the same sense
that the radiation is the mediator of interactions between
electrons, protons and other charged particles), the electron is
the mediator of the interaction between proton and neutron and
acts like a kind of field in the inside of nuclei. Then, we may
solve the [...] equation for the electron to find the form of
interaction between neutron and proton. From the fact that the
electron has the finite rest mass, we may expect that the
interaction energy would decrease rapidly as the distance between
neutron and proton becomes large in comparison with $h/(2\pi
mc)$.
\end{quote}
A critical analysis of Heisenberg's papers on nuclear forces
\cite{HeisenbergEX1,HeisenbergEX2,HeisenbergEX3} led Yukawa to
exclude the picture of the neutron as a bound state made of a
proton and an electron, because of the incompatibility with
quantum mechanics.\footnote{Details can be found in the manuscript
entitled ``The Roles of the Electron for Nuclear Structure'',
collected as document E05 060 U01 in the Yukawa Hall Archival
Library, YITP, Ky\={o}to University, translated in Ref.
\cite{Kawabe1}, pp. 250-252.} In fact, he noticed that the
uncertainty relation would give a value $E = 137 \ mc^2$ for the
neutron mass defect, which differs by a factor $10^2$ from the
correct value $mc^2$ (such a discrepancy had been noticed
also by Heisenberg \cite{HeisenbergEX2}). Likewise, the analysis
of the anomalous dispersion of $\gamma$-rays as due mainly to
neutrons gives the value $42.6 \ mc^2$ for the eigenfrequency of
neutron, which is again very different from that obtained with the
correct mass defect and, moreover, different from the previous
one. Thus Yukawa made the assumption that the neutron is an
elementary particle, and reinterpreted Heisenberg's exchange
interaction between the proton and the neutron, as well as that
between two neutrons, as due to a mediator, the electron, whose
role would be analog to that played by radiation in the
electromagnetic interaction.\footnote{The role of analogies with
quantum electrodynamics in Yukawa's meson theory has been analyzed
by Darrigol in Ref. \cite{Darrigol1}} Following these
guidelines he expected to derive the form of the interaction
between proton and neutron and to explain $\beta$-decay. In
fact, after the recognition that the mediator of the interaction
could not be the electron but a new, much heavier bosonic
particle, he basically followed the same steps outlined in his
1933 talk (as we shall see below). Before trying
to implement his program, Yukawa dwelled on some interesting
considerations concerning the mathematical expression of the
exchange interaction $J(r)$. These are very interesting in view of
understanding how Yukawa arrived at the form $J(r) = - g^2
\frac{e^{- \lambda r}}{r}$, which would later be the basis of his
meson theory. He suggested that this form could be obtained by a
phenomenological approach:

\begin{quote}
To solve this problem, the most effective method at present might
be to calculate the scattering of neutrons on nuclei by assuming a
suitable interaction between neutrons and nuclei, and comparing the
results with experiment (\cite{Kawabe1}, p. 251).
\end{quote}
In particular he hinted to a force decreasing rapidly with the
distance and made reference to a paper by Harrie S. W. Massey
\cite{Massey1}, where elastic collisions of neutrons with
material particles were considered. In this paper the neutron was
modeled as an atom with an electron moving in a field of very high
effective nuclear charge $Z$ (to account for its smallness
with respect to a hydrogen atom), given by
\begin{eqnarray}\label{MasseyPot}
V(r) = e^2 \left(\frac{1}{r} + \frac{Z}{a_0} \right) e^{-
\frac{2Zr}{a_0}},
\end{eqnarray}
$a_0$ being the Bohr radius and ${a_0}/{Z}$ the radius of the
neutron.\footnote{In Ref. \cite{Kawabe1} it is also pointed out
that in a previous paper by Massey \cite{Massey2}, dealing with
the theory of collision of $\alpha$-particles on atomic nuclei, a
calculation of the anomalous scattering of $\alpha$-particles is
carried out by assuming the form $V(r) = - A\frac{ e^{- \mu r}
}{r}$ for the nuclear field near the top of the potential barrier.
According to Kawabe, Yukawa probably was aware also of this paper,
even if he did not quote it in his talk nor in his notes.} This
expression shows a similar behavior if compared with Yukawa's
interaction. Calculations on collisions of neutrons with matter
were also carried out in the same years by Nishina, Tomonaga and
collaborators, who worked also on the neutron-proton interaction
by using various expressions for the potential $J(r)$, including
$A\frac{e^{- \mu r} }{r}$
\cite{TomonagaSpin,NishinaTomonagaTamaki1936}. Furthermore the
topic was the subject of a talk given by Tomonaga, again at the
meeting of the Physico-Mathematical Society of Japan of April 1933
in Sendai, and of a subsequent interplay between Tomonaga and
Yukawa \cite{TomonagaLetter1,Konuma,Yamazaki1}. We will deal in
detail with Tomonaga's contribution in the next Section. One can
infer that in 1933 both Yukawa and Tomonaga had an idea of the
interaction $A\frac{ e^{- \mu r} }{r}$, but the deep meaning of
this formula, and in particular the implications for the range and
the mass of the virtual mediating particle, were still unknown.

After these considerations, Yukawa went back to his theoretical
program, describing it as follows:

\begin{quote}
A neutron can emit an electron and change to a proton, and a
proton can absorb an electron and change to a neutron. This fact
is itself the cause of the interaction between proton and neutron;
in other words, neutron and proton create the electron field,
i.e., the field of the electron wave, and are affected by it
(\cite{Kawabe1}, pp. 251-252).
\end{quote}
In such a situation the total number of electrons changes with
time, so that the electron has to obey to a Dirac equation which
contains a source term involving neutrons and protons. Likewise
protons and neutrons obey a similar Dirac equation with a source
term involving both neutron, proton and electron. These
equations are as follows (Yukawa also included an electromagnetic vector
potential, but he subsequently set it to zero, hence we do not
include it here):\footnote{From this point Yukawa's work is
contained in the manuscript entitled ``Bose Electron'', collected
as document F01 010 U01 in Yukawa Hall Archival Library, YITP,
Kyoto University (\cite{Kawabe1}, pp. 253-257), to which we refer
for details.}
\begin{eqnarray}
\frac{1}{c}L_1&=&\left\{\frac{W}{c}+\rho_1\mathbf{\sigma}\cdot\mathbf{p}+\rho_3mc\right\}\psi=\chi^{\dagger}\gamma\chi\label{EoM1}\, , 
\\
\frac{1}{c}L_2&=&\left\{\frac{W}{c}+\rho'_1\mathbf{\sigma}\cdot\mathbf{p}+\rho_3\left(\frac{1+\tau_3}{2}Mc
+\frac{1-\tau_3}{2}M'c\right)\right\}\chi=\left(\gamma\psi^{\dagger}+\gamma^{\dagger}\psi\right)\chi\label{EoM2} \, ,
\end{eqnarray}
where $\psi$ is the wave function of the electron, $\chi$ is the
wave function of the proton and of the neutron, $L_1$ and $L_2$
are the Dirac operators for electron and neutron-proton
respectively, $\tau_i$ ($i = 1,2,3$) are the isotopic spin matrices
and $\gamma$, $\gamma^{\dagger}$ are two unknown $4 \times 4$
matrices to be determined in order to find the form of the
interaction between electrons, protons and neutrons. Both
equations are derived from the general Lagrangian:
\begin{eqnarray}\label{YukawaLag1}
L = \int \!\!\!\! \int \left[ \psi^{\dagger} L_1 \psi + \chi^{\dagger} L_2
\chi - c \chi^{\dagger} \left( \gamma \psi^{\dagger} +
\gamma^{\dagger} \psi \right)\chi \right] dv dt.
\end{eqnarray}
The $\gamma$ matrices are obtained by requiring that the continuity
equation of electric charge holds on: the final result is
$\gamma = \frac{(\tau_1 + i \tau_2)}{2}\lambda$, $\gamma^{\dagger}
= \frac{(\tau_1 - i \tau_2)}{2}\lambda^{\dagger}$, with $\lambda$
an arbitrary matrix commuting with $\tau_i$.

After having determined the Lagrangian, Yukawa wrote down the
corresponding Hamiltonian and proceeded to the quantization by
imposing Fermi statistics for neutrons, protons and electrons,
that is, anticommutation relations on $\psi$ and $\chi$. From
these, he derived the Heisenberg equations of motion for the
nuclear field operators, which were in perfect agreement with the
previous ones (\ref{EoM2}), while for the electron field operators
the result was different from Eq. (\ref{EoM1}). He noticed that, if Bose statistics (i.e., commutation relations) is
assumed instead for the electron, the results were in agreement with previous
equations of motion.\footnote{Yukawa recalled in an
interview \cite{YukawaAIP} that the idea that electrons within
nuclei had to satisfy Bose-Einstein statistics was suggested to
him by Nishina.} At this point, Yukawa assumed the validity of
equations (\ref{EoM1}) and (\ref{EoM2}) and switched to the task
of finding the specific form of the proton-neutron interaction
from them. He looked for a particular solution of the electron
wave equation satisfying the condition $\psi = 0$ for $\chi = 0$,
obtaining:
\begin{eqnarray}\label{Solution1}
\psi_0 = \left(\frac{p_0}{c} - \rho_1 \mathbf{\sigma} \cdot
\mathbf{p} - \rho_3 mc  \right) \! \frac{1}{4 \pi \hbar^2} \!\!\!\!
\bigints \!\!\!\!\!\!\!\!\!\! \bigints \!\!\!\!\!\!\!\!\!\! \bigints \!\!\!\! \frac{\chi^{\dagger} \gamma
e^{\frac{i}{\hbar} \rho_3 mc \left| {\bf r}^{'} - {\bf r} \right|
} \chi \! \left( {\bf r}^{'}, t- \frac{\left| {\bf r}^{'} - {\bf r}
\right|}{c} \right) }{\left| {\bf r}^{'} - {\bf r} \right|} dv^{'} \!
+ \! ... \, .
\end{eqnarray}
Then, upon substituting this solution into the Hamiltonian, he
finally obtained the interaction energy, which however had a
Coulomb form, not the expected behavior. He concluded that a
well behaved result could be obtained by taking into account the
Fermi statistics for the electron, but the calculations appeared to
be too hard to deal with. This failure was acknowledged in
the talk,\footnote{The presumed text of the talk is the
manuscript entitled ``A Comment on the Problem of Electrons in the
Nucleus'', collected as document E05 080 U01 in the Yukawa Hall
Archival Library, YITP, Kyoto University, translated in
\cite{Kawabe1}, pp. 248-249.} where Yukawa admitted that

\begin{quote}
In any case, the practical calculation does not yield the
looked-for result that the interaction term decreases rapidly as
the distance becomes larger than $h/(2\pi
mc),$\footnote{Interestingly, neither here nor in the
abstract, a mention is made of the fact that a range determined
by the Compton wavelength of the electron, $h/(2\pi mc)$, would
have been in any case too large.} unlike what I wrote in the
abstract of this talk (\cite{Kawabe1}, p.249).
\end{quote}
In Yukawa's opinion, the failure had its roots in the
incompleteness of relativistic quantum mechanics, which led him to
study quantum field theoretical issues and, in particular,
foundational problems in the following year. Indeed, the
1933 talk begins right away with the words:

\begin{quote}
The problems of atomic nuclei, especially the problem of electrons
in the nucleus, might not be solved until we reflect on the
foundation of quantum mechanics and complete the correct
relativistic quantum theory. It cannot be solved only by partial
and formal modification of the theory (\cite{Kawabe1}, p. 248).
\end{quote}
Accordingly, in April 1934, again at the annual meeting of the
Physico-Mathematical Society of Japan, he gave the talk ``On the
Probability Amplitude in Relativistic Quantum Mechanics'',
focusing on topics related to the foundations of quantum field
theory, where an early attempt at a better foundation for
relativistic quantum mechanics was reported. In this talk he
introduced for the first time his idea of the {\it maru}, which
literally means ``circle'', but in fact denotes a closed surface
within the four dimensional space-time on which fields
should be quantized.\footnote{See the manuscript entitled ``On
Probability Amplitudes in Relativistic Quantum Mechanics'',
collected as document F01 030 T02 in the Yukawa Hall Archival
Library, YITP, Kyoto University (\cite{Kawabe1}, pp. 258-261).}
This work was a precursor of the non-local field theory he would
have developed in 1950s, while it later inspired Tomonaga in
developing his super-many-time theory \cite{TomonagaSMT}, which
formed the basis of his Nobel prize winning approach to covariant
QED. Yukawa strongly believed that a solution of the internal
inconsistency problem in relativistic quantum field theory would
provide useful insights for the construction of an unifying
picture of elementary particles.

\subsection{Yukawa's theory of nuclear forces and the meson}

There were no novelties in Yukawa's research on nuclear forces
until late spring 1934, when Yukawa read Fermi's paper on
$\beta$-decay and became aware of Pauli's neutrino hypothesis.
Then he independently tried to construct a model of nuclear force
based on the exchange of the pair electron-neutrino but, as
Heisenberg and other physicists before him, he found that the
resulting force was too small to be a viable candidate. So he
changed his strategy. Instead of trying to get the nuclear
forces from the existing theory, he started focusing on the
characteristics of the nuclear force field which were inferred
from experiment. He postulated the existence of a
brand-new field which would be in the same relation to nuclear
forces as the electromagnetic field is to electrostatic forces,
and was tailored to meet the phenomenological requirements. His
struggle ended in October, when the fog disappeared and the truth
arose:

\begin{quote}
The nuclear force is effective at extremely small distances, on
the order of 0.02 trillionth of a centimeter. That much I knew
already. My new insight was the realization that this distance and
the mass of the new particle that I was seeking are inversely
related to each other. Why had I not noticed that before? The next
morning, I tackled the problem of the mass of the new particle and
found it to be about two hundred times that of the electron. It
also had to have the charge of plus or minus that of the electron.
Such a particle had not, of course, been found, so I asked myself,
``Why not?'' The answer was simple: an energy of 100 million
electron volts would be needed to create such a particle, and
there was no accelerator, at that time, with that much energy
available (\cite{Tabibito}, p. 324).
\end{quote}
The work was presented in November 1934 to the meeting of the
T\={o}ky\={o} Center of the Physico-Mathematical Society of Japan
and met the interest of Nishina.\footnote{This circumstance is
clarified by Kawabe \cite{Kawabe2}. Indeed in his autobiography
\cite{Tabibito} Yukawa mentions a talk given at the regular
meeting of the \={O}saka branch of the Physico-Mathematical
Society of Japan and then a second talk given in November in
Tokyo. The passage including this talk in T\={o}ky\={o} is missing
in the English translation by L. Brown and R. Yoshida.
Furthermore, as pointed out by Kawabe \cite{Kawabe2} on the basis
of archival sources, the \={O}saka branch of the
Physico-Mathematical Society of Japan started only on June 1,
1935. The T\={o}ky\={o} talk is also quoted by Kobayashi in his
recollections \cite{KobayasiM}.} The corresponding paper, Yukawa's
famous meson paper, was written and submitted for publication by
the end of the same month.

Before addressing Yukawa's work, it is very useful to quote a
passage of his first publication, i.e. the introduction to the
Japanese translation of the 1932 Heisenberg's papers on nuclear
structure, written in 1933 \cite{Brown2}:

\begin{quote}
Though Heisenberg does not present a definite view on whether
neutrons should be seen as separate entities or as combinations of
a proton and an electron, this problem, like the $\beta$ decay
problem stated above, cannot be resolved with today's theory. And
unless these problems are resolved, one cannot say whether the
view that electrons have no independent existence in the nucleus
is correct (\cite{Brown2}, p. 122).
\end{quote}
As testified by his words, the starting point of his investigation
is a deep analysis of Heisenberg's picture of neutrons and of
unsolved problems, like $\beta$ decay. Indeed Fermi's theory would
have been published a year later, in 1934, and would have been
taken as a reference in his 1935 seminal work \cite{Yukawa1935},
together with the {\it Platzwechsel} interaction postulated by
Heisenberg and the conclusions by Tamm \cite{Tamm1} and Iwanenko
\cite{Iwanenko1}, which led to rule out a
neutron-proton interaction mediated by an exchange of an electron-neutrino
pair.\footnote{The content of Yukawa's first paper on nuclear
forces had already been anticipated in a talk given at the meeting
of the Physico-Mathematical Society of Japan, held at Tokyo
University in November, 1934. But, as Kobayashi reported in his
recollections, ``his hypothesis was too bold to be in line with
common sense and, at best, gave one the impression of being
somewhat interesting if the proposed particle indeed existed''
(\cite{KobayasiM}, p.7). Yukawa also submitted an article with an
outline of his proposal to {\it Physical Review} but it was
rejected.} Then he clearly outlines his strategy and the ultimate
goal of his work:

\begin{quote}
To remove this defect, it seems natural to modify the theory of
Heisenberg and Fermi in the following way. The transition of a
heavy particle from neutron state to proton state is not always
accompanied by the emission of light particles, i.e., a neutrino
and an electron, but the energy liberated by the transition is
taken up sometimes by another heavy particle, which in turn will
be transformed from proton state into neutron state. If the
probability of occurrence of the latter process is much larger
than that of the former, the interaction between the neutron and
the proton will be much larger than in the case of Fermi, whereas
the probability of emission of light particles is not affected
essentially. Now such interaction between the elementary particles
can be described by means of a field of force, just as the
interaction between the charged particles is described by the
electromagnetic field. The above considerations show that the
interaction of heavy particles with this field is much larger than
that of light particles with it (\cite{Yukawa1935}, p. 48).
\end{quote}
Also here the analogy with the electromagnetic case is pursued, so
that Yukawa's purpose is to look for this new field, to define its
quantum and to investigate the corresponding properties. As a
starting point, he adopts a semiclassical approach. The required
scalar field $U(x,y,z,t)$ has to decrease rapidly with the
distance, so its behavior is assumed to be described by the
following function:
\begin{eqnarray}\label{YukawaU}
\pm g^2 \, \frac{e^{- \lambda r}}{r},
\end{eqnarray}
which is the static, spherically symmetric solution of the
Klein-Gordon equation in vacuum:
\begin{eqnarray}\label{KleinGordon1}
\left( \Delta - \frac{1}{c^2} \frac{\partial^2}{\partial t^2} -
\lambda^2 \right) U = 0.
\end{eqnarray}
In Eq. (\ref{YukawaU}), $g$ and $\lambda$ are constants with the
dimensions of an electric charge and of cm$^{-1}$, respectively, so
that the range of forces is of the order of ${1}/{\lambda}$.

Then Yukawa adds heavy particles (i.e., neutrons and protons) in
order to study their interaction with the $U$-field, whose effect
is the transition from the neutron state to the proton state. This
new situation is implemented by adding a source term to Eq.
(\ref{KleinGordon1}), so that the $U$-field satisfies the new
equation:
\begin{eqnarray}\label{KleinGordon2}
\left( \Delta - \frac{1}{c^2} \frac{\partial^2}{\partial t^2} -
\lambda^2 \right) U = - 4 \pi g \, \tilde{\psi}
\left(\frac{\rho^{\xi} - i\rho^{\eta}}{2} \right) \psi,
\end{eqnarray}
$\rho^{\xi}$ and $\rho^{\eta}$ being the isotopic spin matrices
introduced in Eq. (\ref{matrix}) and $\psi$ is the wave function
of the heavy particles (which is a function of time, position,
spin and isotopic spin).

Likewise, the complex conjugate field $\tilde{U}(x,y,z,t)$ is
introduced, satisfying the equation
\begin{eqnarray}\label{KleinGordon3}
\left( \Delta - \frac{1}{c^2} \frac{\partial^2}{\partial t^2} -
\lambda^2 \right) \tilde{U} = - 4 \pi g \, \tilde{\psi}
\left(\frac{\rho^{\xi} + i\rho^{\eta}}{2} \right) \psi,
\end{eqnarray}
and gives rise to the inverse transition from the proton state to
the neutron state. In pursuing the analogy with the
electromagnetic case, at this point Yukawa does not take into
account the {\it vector} potential ``as there's no correct
relativistic theory for the heavy particles'' (\cite{Yukawa1935},
p. 50). He would have worked out the complete vector meson theory
only later, in 1938 \cite{Yukawa3}.  Then, from a non relativistic
wave equation for the heavy particle, by neglecting the spin and a
constant term, he was able to derive the full Hamiltonian for a
system of two heavy particles at positions $(x_1,y_1,z_1)$ and
$(x_2,y_2,z_2)$ and with a small relative velocity:
\begin{eqnarray}\label{YukawaHamiltonian}
H = \frac{{\bf p}_1^2}{2M} + \frac{{\bf p}_2^2}{2M} + \frac{g^2}{2} \left( \rho_1^{\xi} \rho_2^{\xi} + \rho_1^{\eta} \rho_2^{\eta} \right) \frac{e^{- \lambda r_{12}}}{r_{12}} + \left(  \rho_1^{\zeta} +  \rho_2^{\zeta} \right) D,
\end{eqnarray}
where ${\bf p}_1$, ${\bf p}_2$ are the momenta of the heavy
particles, $(\rho_1^{\xi}, \rho_1^{\eta}, \rho_1^{\zeta})$ and
$(\rho_2^{\xi}, \rho_2^{\eta}, \rho_2^{\zeta})$ are the
corresponding isotopic spin matrices, $r_{12}$ is the distance
between them and $D$ is the mass defect between proton and
neutron. As he pointed out:\footnote{Notice that here Yukawa
identifies $J(r)$ with Heisenberg's ``Platzwechseintegral'', as
noted by Carson \cite{Carson2}, as well as Miller
\cite{MillerBook1994}, and interprets it as a real migration of a
Bose particle, which will be soon identified with the quantum of
his $U$-field.}

\begin{quote}
This Hamiltonian is equivalent to Heisenberg's Hamiltonian, if we
take for ``Platzwechseintegral'' $J(r) = - g^2 \frac{e^{- \lambda
r}}{r}$, except that the interaction between the neutrons and the
electrostatic repulsion between the protons are not taken into
account (\cite{Yukawa1935}, p. 51).
\end{quote}
Indeed a comparison with Heisenberg's expression in Eq.
(\ref{Hamiltonian1}) clearly shows analogies and differences.
Besides the explicit expression of $J(r)$, its overall sign comes
out negative, so that the lowest energy state of $H^2$ has spin
$1$ as required by experiments. This happens at odds with
Heisenberg's choice, which brought to a lowest energy
state with spin $0$. Finally, the constants $g$ and $\lambda$ can
be determined from experimental data, and an agreement is found
when taking $\lambda$ between $10^{12}$ cm$^{-1}$ and $10^{13}$
cm$^{-1}$ and $g$ as a multiple of the elementary charge $e$. Here
Yukawa briefly hints to a possible derivation of the range of the
interaction between neutron and proton, starting from the
calculation of the mass defect of $H^2$ and the probability of the neutron-proton 
scattering under the assumption that the relative
velocity is small with respect to the speed of light. In this
respect, in a footnote he very interestingly mentions similar
calculations carried out by Tomonaga:

\begin{quote}
These calculations were made previously, according to the theory
of Heisenberg, by Mr. Tomonaga, to whom the writer owes much. A
little modification is necessary in our case. Detailed accounts
will be made in the next paper (\cite{Yukawa1935}, p. 52).
\end{quote}
From archival documents
\cite{Archive1,YukawaArchive,TomonagaLetter1,Konuma} and
recollections by Tomonaga \cite{TomonagaSpin}, the role he played
in the early stages of the birth of meson's theory, which is
hinted at in this footnote, clearly emerges \cite{Yamazaki1}. We
will come to this issue in the following Section.

Yukawa's next step is to determine the nature of the quanta that
the general principles of quantum field theory associate to the
$U$-field (which would then play for this field the same
role that photons play for the electromagnetic field), by taking
into account the fact that they should follow Bose statistics and
have a charge equal to $+e$ or $-e$. As such, the operator
associated to $U$ works by decreasing by one the number of quanta
with negative charge, or increasing by one the number of quanta
with positive charge, and satisfies the relativistic wave equation:
\begin{eqnarray}\label{waveEq1}
\left( p_x^2 + p_y^2 + p_z^2 - \frac{W^2}{c^2} + \lambda^2 \right)
U = 0,
\end{eqnarray}
which can be obtained from Eq. (\ref{KleinGordon1}) by identifying
$p_x = -ih \frac{\partial}{\partial x}$, $p_y = -ih
\frac{\partial}{\partial y}$, $p_z = -ih \frac{\partial}{\partial
z}$, $W = ih \frac{\partial}{\partial t}$. This equation shows
that the quantum associated to the $U$-field has
-- unlike the photon -- a finite mass $m_U = {\lambda
h}/{c}$, which is related to the finite range of the nuclear force.
By substituting $\lambda = 5 \cdot 10^{12}$ cm$^{-1}$, an estimate
of $m_U$ is obtained which is about $200$ times the electron
mass.\footnote{Later Gian Carlo Wick found that the relation
between the mass of the mediator of the nuclear force and the
range of this force in Yukawa's theory could be understood simply as 
a consequence of Heisenberg's uncertainty relations, rather
than a result of perturbation theory \cite{Wick1}.} The new
particle was called {\it heavy quantum}\footnote{As
opposed to {\it light} quantum, i.e., electrons and neutrinos
which had been previously thought to be associated with the
nuclear force. Notice the possible pun here, related to the double
meaning of the word ``light''.} while the term {\it meson} was
coined by Homi Bhabha in 1939 to point out that its mass is
intermediate between the electron and the proton masses
\cite{Bhabha}. A comment is made on the possible observation of
this massive particle in nuclear transformations, and Yukawa shows
how the energies required for its production are not available in
ordinary nuclear reactions. In fact, he shows that with
typical nuclear energies, the wave function associated to $U$-quanta is an evanescent function, leading to a negligible
probability of observing a $U$-quantum outside the
nucleus.\footnote{At the time the concept of virtual particles
and of their role in interactions was not clear yet
\cite{Carson1,Carson2}.} In his concluding remarks at the end
of the paper, he suggests that the predicted massive quanta ``may
have some bearing on the shower produced by cosmic rays''
(\cite{Yukawa1935}, p. 57). Indeed, many
particles with such a huge energy could be found within cosmic rays, and a year later,
in 1936, a particle compatible with that predicted by Yukawa was
actually discovered in cosmic ray showers \cite{AndersonNed1936}.
This particle was later identified with the {\it muon}
while the true meson (the {\it pion}) would have discovered only
in 1947 \cite{Lattes}. In fact, with his work, Yukawa began to go
beyond the domain of nuclear physics, paving the way towards the
regime of high energy physics. The particle physics era was about
to start, as noticed by Brown \cite{Brown2}.

The final part of Yukawa's seminal work is devoted to an
alternative formulation of $\beta$-decay theory in which the
quanta of the $U$-field are assumed to interact with a light
particle, which jumps from a neutrino state of negative energy to
an electron state of positive energy. This process is implemented
by making the $U$-field to interact with an electron $\Psi_k$ and
a neutrino $\varphi_k$ field ($k = 1, 2, 3, 4$), which amounts to
add the new source term $- 4 \pi g^{'} \sum_{k = 1}^4
\tilde{\Psi}_k \varphi_k$ to the right hand side of Eq.
(\ref{KleinGordon3}). Here $g^{'} $ is a coupling constant with
the same dimension as $g$. A comparison between the matrix element
corresponding to the above process and Fermi's result shows that
there is perfect agreement upon identifying $\frac{4 \pi g
g^{'}}{\lambda^2}$ with Fermi's $g$ constant, whose value is $4
\cdot 10^{-50}$ cm$^3$ erg. By substituting $\lambda = 5 \cdot
10^{12}$ and $g = 2 \cdot 10^{-9}$, Yukawa found $g^{'} = 4 \cdot
10^{-17}$, which is smaller than $g$ by a factor $10^{-8}$. So his
final comment is:

\begin{quote}
This means that the interaction between the neutrino and the
electron is much smaller than that between the neutron and the
proton so that the neutrino will be far more penetrating than the
neutron and consequently more difficult to observe
(\cite{Yukawa1935}, p. 56).
\end{quote}
For the first time a clear difference emerges between
the weak and the strong nuclear force: a qualifying feature of
Yukawa's theory is that it provides a double mechanism able to
explain the force between neutrons and protons, via the exchange of a
meson, which carries energy, momentum and electric charge, as well
as nuclear $\beta$-decay, via the decay with small probability of an
electrically negative (positive) meson into an
electron-antineutrino (positron-neutrino) pair. Notice,
moreover, that this view is very similar to the modern
understanding of weak interactions as mediated by intermediate
vector bosons.

Yukawa's work was quite ignored abroad for a few years, given the fact that
western scientists were not inclined to accept the idea of a new
particle without experimental evidence. This is testified, for instance,
by Pauli's reluctance to publicly mention his neutrino's
hypothesis \cite{PaisIWB,PauliNeutrino1,PauliSolvay1933} or by
Bohr's negative attitude towards Yukawa's results, clearly
expressed during his visit to Japan in the spring of 1937
\cite{Darrigol1,Spradley}. To this respect, a comment by Tomonaga
is enlightening:

\begin{quote}
It seems to me [...] that the reason for this rejection is
that although one wall after another was being removed and although
there were many new discoveries, there remained the stubborn
prejudice that there was a sanctuary inside nuclei, and the
physics community was allergic to new particles. I might note that
the workplaces of Heisenberg, Fermi and Yukawa, who successively
removed the walls of the sanctuary, get farther and farther from
Copenhagen where Bohr resided. This might mean that Bohr's
influence gets weaker as you move farther away
(\cite{TomonagaSpin}, p. 183).
\end{quote}

These circumstances probably concurred to determine the
successful endeavor of Yukawa, rather than that of some European
physicist, despite the principles of new quantum mechanics had
begun to spread in Japan only in 1929, after Nishina's return from
Europe \cite{Brown2}. Only after the discovery of a new charged
particle in cosmic ray showers by Anderson and Neddermeyer
\cite{AndersonNed1936}, whose mass was compatible with meson's
mass, there appeared the first reference to Yukawa's paper outside
Japan, in a work by Oppenheimer and Serber \cite{OppSerber}. A
paper was then submitted by Yukawa \cite{YukawaCosmic},where 
he suggested the identification of his meson with the new
particle discovered in cosmic rays. This prompted him to resume
his theory and to develop it in a series of papers with various
collaborators \cite{Yukawa3,Yukawa2,Yukawa4} until 1938, when his
idea finally began to find wider acceptance.

\section{Tomonaga and the neutron-proton interactions}

The aim of this Section is to reconstruct the role played by
Tomonaga Shin'ichir\={o} in the birth of meson theory, by analyzing archival documents 
as well as Tomonaga's recollections \cite{TomonagaSpin}. Some further biographical information about Tomonaga is
included as well.

A useful starting point is Lecture 12 of Tomonaga's book
\cite{TomonagaSpin}, referring to his time as third year
undergraduate student and Yukawa's classmate at Ky\={o}to
University, as well as to the beginning of his career as a
scientist in Nishina's Lab at RIKEN. Indeed, as already remarked, for the young Tomonaga (as well as
for Yukawa) the undergraduate period marked the beginning of a deep interest in quantum mechanics.
The reading of Heisenberg's quantum resonance paper
\cite{Heisenberg1} strongly impressed him:

\begin{quote}
[...] I think I was attracted to this paper more because of
Heisenberg's expert use of analogy. I was attracted by the
deftness of the analogy in which he started from the resonance of
two pendula, which is a very ordinary, everyday phenomenon, and
gradually proceeded to the sophisticated problem of the symmetry
of $\psi$ and the statistics of the particle (\cite{TomonagaSpin},
pp. 221-222).
\end{quote}
Heisenberg's and Dirac's 1929 lectures in Japan were, however, even more significant for his scientific growth. 

\begin{quote}
Miraculously, I remember, I could more or less understand the
content of the lectures because fortunately I had already looked
through papers related to these talks. [...] This was the first
time I had come from rural Kyoto to Tokyo and seen in person
distinguished people like Professor Hantaro Nagaoka, Professor
Nishina, and Professor Sugiura and also the brilliant graduates of
the University of Tokyo, who obviously looked very bright. I
listened to the lectures, hiding myself toward the last row of the
room, overwhelmed by those luminaries (\cite{TomonagaSpin}, p.
222).
\end{quote}
A further intensive series of lectures on quantum mechanics took
place in Ky\={o}to at the beginning of 1930, organized by the
spectroscopist Kimura Masamichi, who also recognized the relevance
of theoretical physics, with particular regard to quantum physics, after
visiting Europe and America. The lectures were given by Sugiura
Yoshikatsu\footnote{Sugiura's lectures began on 13 January 1930
and went on for about a month at a pace of three lectures per
week. The focus was mainly on applications of quantum mechanics,
even though he made a quick introduction to general concepts and
framework, such as matrix mechanics, wave mechanics, $q$-numbers
and group theory. Then he dealt with the study of periodic and
non-periodic systems, ending with a discussion of the molecule
problem and further applications of quantum mechanics to
chemistry. See Ref. \cite{Sugiura} and references therein for
further details on his lectures, his scientific trajectory and his
role in spreading quantum mechanics ideas in Japan.} of RIKEN,
while in the early summer of the subsequent year also Nishina came
to lecture in Ky\={o}to. The style of lecturing of Nishina was
very different from Sugiura's one, as vividly recalled by
Tomonaga, who was deeply impressed:

\begin{quote}
Heisenberg's book {\it Physikalische Prinzipien der
Quantentheorie} (The physical principles of quantum theory) was
used as the text for Professor Nishina's lectures. When Professor
Sugiura would lecture, he would write a long, long formula [...]
from one end to the other of a long blackboard, and he would
lecture about his own work. It might have been a creative work,
but it was too detailed for a beginner to make sense out of it. On
the other, Professor Nishina's lecture, albeit much of it was from
the book used and much credit should go to Heisenberg, was
impressive, especially the discussions after the lectures
(\cite{TomonagaSpin}, p. 226).
\end{quote}
Tomonaga's scientific career would have been significantly shaped
by Nishina who, in 1932, invited him to carry out research in his
lab at RIKEN. Chadwick's discovery of the neutron
\cite{Chad1932,Chad1932a} and Heisenberg's subsequent papers on
nuclear theory \cite{HeisenbergEX1,HeisenbergEX2,HeisenbergEX3}
drew Nishina and Tomonaga interest towards the investigation of
the properties of nuclear forces. In this context the deuteron, as
a simple two-body system, soon appeared the ideal candidate to
work out:

\begin{quote}
[...] I started calculations related to phenomena such as the
binding energy of the deuteron and the scattering or capture of a
neutron by a proton. Heisenberg regarded the nuclear force as an
exchange force and introduced the potential $J(r)$ for it. He
figured that the nuclear force must act only over a very short
range and that the potential goes to zero if $r \ge 10^{13}$ cm.
Now, it was necessary to determine the magnitude of the potential.
Since the binding energy of the deuteron was known experimentally,
it was possible to determine the magnitude of the nuclear force so
that the theoretical value for the binding energy agreed with
experiment. Using the potential, we can discuss neutron scattering
and capture. As more and more experimental facts surfaced, both
the cross-section of elastic collision and that of the capture of
a neutron by a proton were found to be abnormally large for slow
neutrons, and this drew Professor Nishina's attention
(\cite{TomonagaSpin}, pp. 227-228).
\end{quote}
Unfortunately, neutron-proton scattering calculations based on the
potential $J(r)$, previously extracted from the binding energy of
the deuteron, did not agree with experimental findings. Tomonaga
carried out calculations by assuming various forms for the
short-range interaction $J(r)$, among which we find $A \frac{e^{- \lambda
r}}{r}$, which coincides with the one later introduced by Yukawa,
but the conclusions did not change. Indeed, he found that a very
large scattering cross-section could be obtained only by
postulating the existence of a novel $S$ state of the deuteron
with zero energy, in addition to the usual one. However, in order to
get this additional level, Majorana exchange force had to be added
to Heisenberg's force,\footnote{Notice that here Tomonaga is
clearly referring to Majorana's work on exchange forces
\cite{Majorana1}. Interestingly, he mentions Majorana also in the
letter addressed to Yukawa \cite{TomonagaLetter1} after the 1933
Sendai talk. Thus we can conclude that Tomonaga was aware of
Majorana's results. We may guess that also Yukawa was aware of
this work as well, as in fact is claimed by Brown (cf.
\cite{Brown2}, p. 97), but he never mentions Majorana's paper in
his celebrated meson article \cite{Yukawa1935}. He will hint at
Majorana' exchange force only later, in his Nobel Lecture
\cite{YukawaNobel}. We are grateful to Francesco Guerra for
pointing out to us the latter point.} Tomonaga providing also the
ratio of the forces, and obtaining a good agreement with experimental
results. Nishina and Tomonaga reported on their calculations at
the spring meeting of the Physico-Mathematical Society of Japan
held in April 1933 in Sendai \cite{TomonagaSpin}\cite{Yamazaki1}
(see also subsection 6.2); the abstract of their talk
entitled ``Scattering of neutron by proton'' reads:

\begin{quote}
We have analyzed scattering of neutron by proton using
Heisenberg's theory on nuclear structure, assuming the shape of
interactions between neutron and proton, and taking into account
the mass defect of hydrogen 2. Our result has been compared with
experimental results (\cite{Konuma}, p. 013009-2).
\end{quote}
At Yukawa's request, a detailed seven page letter of Tomonaga
followed (probably written in May or June 1933), in which he gave
further details on his scattering calculations
\cite{TomonagaLetter1,Yamazaki1}. Also here, various forms of
$J(r)$ are mentioned, including $A \frac{e^{- \lambda r}}{r}$, and
an estimate of the corresponding range of interaction $\lambda$ is
provided by fitting the experimental data with the theoretical curve,
giving the value $7 \cdot 10^{12}$ cm$^{-1}$. A very
interesting interplay between the two young physicists started, as
testified by a number of letters, in addition to the one quoted above
\cite{YukawaArchive,Kawabe1,Konuma}. Tomonaga's own
interesting results were not promptly published:

\begin{quote}
I was quite elated with this achievement, and Professor Nishina
was also satisfied, and our results were reported [...] in Sendai,
1933 [...]. However, probably because he was so busy with a
variety of experimental work, Professor Nishina put off publishing
this paper. While I was agonizing about this, Bethe and Peierls
did exactly the same thing and published it. I was extremely
upset, and I was livid with Professor Nishina (\cite{TomonagaSpin}, p. 228).
\end{quote}
Here Tomonaga referred to analogous calculations carried
out and published by Bethe and Peierls in 1935
\cite{BetheP1935a,BetheP1935b}, while his paper would have been
published only later, in 1936 \cite{NishinaTomonagaTamaki1936}.

In the same 1933 letter to Yukawa, Tomonaga also included the
results of a calculation on the neutron capture by the proton. In
such a case, the same strategy adopted for the scattering problem
did not work, because the transition involved should be $P
\rightarrow S$ and the assumption of the additional $S$ level had
no influence on the $P$ wave. A possible solution would have been to
assume a further $P$ level very close to zero energy, but a price
had to be paid by changing the mathematical expression of $J(r)$, thus loosing
the agreement with experimental results \cite{TomonagaSpin}. Here, the following potential forms were
adopted \cite{YukawaArchive,Kawabe1},
\begin{eqnarray}\label{Capture1}
\frac{P e^{- \lambda r}}{1-e^{- \lambda r}}, \  \  \frac{P}{(1+e^{ \lambda r})(1+e^{- \lambda r})} \, ,
\end{eqnarray}
because the radial Schr\"{o}dinger equation was solvable for $\ell =
0$. The problem would have been later solved by Fermi by taking
into account also the magnetic moments of the neutron and the
proton, and allowing emission via magnetic dipole in addition to
the usual one via electric dipole when the capture takes place.

In summary, during the years 1933-35 also Tomonaga and Nishina
were working on the neutron-proton force and its range, and,
interestingly, in their calculations they adopted various
mathematical forms for the short range interaction potential
$J(r)$, including Yukawa's one.

\section{Concluding remarks}

In this paper we focused on the development of the concept of
exchange forces in the realm of nuclear physics, starting from
Heisenberg's pioneering papers. Then we analyzed Majorana's
decisive improvements to Heisenberg's theory, as well as Fermi's theory of
$\beta$-decay, which proved to be crucial intermediate steps towards the idea of a
force mediated by virtual quanta, the last step in this path
being Yukawa's meson theory, whose genesis has been carefully
reconstructed. Fermi's and Yukawa's theories were the first
quantum field theories after quantum electrodynamics to be
established.

The relevance of Japanese contributions to nuclear as well as
particle physics can hardly be underestimated, and this conclusion applies as well to
the rapidity of assimilation of European science. Indeed, starting from the 1920s and throughout the 1930s, 
Japanese physicists became a consolidated
presence in European universities, so that they could absorb there the
new developments, contribute to them and bring them back to their
country. It was Nishina who, back in Japan in 1929 after a
eight-years research stay in Europe, brought in his country the
``Copenhagen spirit'', and with it the new quantum mechanics. He
strongly contributed to the spreading of the new ideas and to the
engagement of the new generation of physicists, also by inviting
distinguished European scientists in Japan, such as Heisenberg and
Dirac. Young Japanese physicists were able to go beyond the common
thinking, thus making the decisive steps toward a quantum field
theory of the nuclear interactions just by postulating the existence of new particles.
Yukawa and Tomonaga were among them; they both were deeply impressed
by the new quantum mechanics, and both gave decisive contributions to
nuclear theory. Yukawa's key assumptions were that the principles
of quantum theory did apply inside the nucleus, and that the
nuclear interaction was a fundamental force just as
electromagnetism. His main achievement was the introduction of a
mediating virtual particle for nuclear forces, also establishing the inverse
proportionality between the range of a force and the mass of its
mediator, but he finally conceived also the actual distinction between the two nuclear forces 
-- the strong and weak nuclear interactions -- with different
coupling constants. All these concepts would have reached full
maturity only after World War II, but Yukawa's theory was
undoubtedly a decisive step towards it. In the course of these developments,
a fruitful interplay between Yukawa and Tomonaga
emerged in 1933-34, as testified by various sources, hinting to a role
played by Tomonaga in the birth of meson theory that is
interesting, and certainly deserves further investigation.

\backmatter


\end{document}